\documentclass[reqno,12pt]{amsart}
\usepackage{fullpage}
\usepackage{amsfonts}
\usepackage{amssymb}

\numberwithin{equation}{section}
\def\X{{\bf X}}
\def\Y{\hat{\bf X}}
\def\pr{{\rm pr}}
\def\p{\partial}
\def\Div{{\rm Div}}
\def\lieder#1{{\mathcal L}_{#1}}
\def\til{\tilde}

\def\Esp{{\mathcal E}}

\def\triv{\text{triv.}}
\def\ext{\text{ext.}}
\def\tbinom#1#2{{\textstyle\binom{#1}{#2}}}

\def\const{{\rm const.}}
\def\Rnum{\mathbb{R}}

\def\arb{{\rm arb.}}
\def\tint{{\textstyle\int}}

\def\parder#1#2{\frac{\partial{#1}}{\partial{#2}}}
\def\parderop#1{\partial/\partial{#1}}
\def\extder#1{\mathbf{d}{#1}}

\newtheorem{prop}{Proposition}
\newtheorem{thm}{Theorem}
\newtheorem{cor}{Corollary}

\def\Ref#1{Ref.\cite{#1}}

\def\ie/{i.e.}
\def\cf/{cf.}
\def\eg/{e.g.}
\def\etc/{etc.}

\begin{document}
\allowdisplaybreaks[3]

\title{On the incompleteness of Ibragimov's conservation law\\theorem and its equivalence to a standard formula using symmetries and adjoint-symmetries}

\author{
Stephen C. Anco
\\\lowercase{\scshape{
Department of Mathematics and Statistics\\
Brock University\\
St. Catharines, ON L2S3A1, Canada}} \\
}

\begin{abstract}
A conservation law theorem stated by N.\ Ibragimov
along with its subsequent extensions
are shown to be a special case of a standard formula
that uses a pair consisting of a symmetry and an adjoint-symmetry
to produce a conservation law through a well-known Fr\'echet  derivative identity.
Also, the connection of this formula (and of Ibragimov's theorem)
to the standard action of symmetries on conservation laws is explained,
which accounts for a number of major drawbacks that have appeared
in recent work using the formula to generate conservation laws.
In particular, the formula can generate trivial conservation laws
and does not always yield all non-trivial conservation laws
unless the symmetry action on the set of these conservation laws is transitive.
It is emphasized that all local conservation laws
for any given system of differential equations
can be found instead by a general method using adjoint-symmetries.
This general method is a kind of adjoint version of the standard Lie method
to find all local symmetries and is completely algorithmic.
The relationship between this method, Noether's theorem,
and the symmetry/adjoint-symmetry formula is discussed.
\end{abstract}

\maketitle

\section{Introduction}
\label{intro}

The most well-known method for finding conservation laws of differential equations (DEs)
is Noether's theorem \cite{Noe}
which is applicable to any system of one or more DEs
admitting a variational formulation in terms of a Lagrangian.
Noether's theorem shows that every local symmetry preserving
the variational principle of a given variational system
yields a non-trivial local conservation law.
Moreover, for variational systems that do not possess any differential identities,
every non-trivial local conservation law arises from some local symmetry that
preserves the variational principle.

However,
there are many physically and mathematically interesting DEs that are not variational systems,
and this situation has motivated a lot of work in the past few decades to look for
some generalization of Noether's theorem which could be applied to non-variational DEs.
One direction of work has been to replace the need for a variational principle
by introducing some other structure,
but still making use of the local symmetries of a given DE system
to produce local conservation laws.
In fact,
a general formula is available that yields local conservation laws
from local symmetries combined with solutions of the adjoint of the symmetry determining equations.
This formula first appears (to the knowledge of the author)
in a 1986 paper by G.\ Caviglia \cite{Cav}
and was later derived independently in a 1990 Russian paper by F. Lunev \cite{Lun}
as well as in a 1997 paper by the author and G.\ Bluman \cite{AncBlu97}. 
In the latter paper, 
solutions of the adjoint of the symmetry determining equations were called adjoint-symmetries; 
these solutions are also known as cosymmetries 
in the literature on integrable systems \cite{Bla}. 
Essentially the same formula appears
in a more abstract form in the cohomological framework for finding conservation laws,
summarized in \Ref{KraVin,Ver,Zha86}.

In recent years,
a similar conservation law formula has been popularized by N.\ Ibragimov
\cite{Ibr07,Ibr07b,Ibr10,Ibr11},
and subsequently extended by others \cite{Gan11,Gan14,Zha,ZhaXie},
where a ``nonlinear self-adjointness'' condition is required to hold for the given DE system.
However, in several papers \cite{Fre13,Gan15,ZhaXie},
this formula sometimes is seen to produce only trivial conservation laws,
and sometimes the formula does not produce all admitted conservation laws.
Also, in a number of papers \cite{IbrTorTra10,IbrTorTra11,Gan11,FreSam12,Gan14,Zha,ZhaXie},
the use of translation symmetries is mysteriously avoided
and other more complicated symmetries are used instead, without explanation.

The purpose of the present paper is to make several relevant remarks:
\newline
(1) Ibragimov's conservation law formula is a simple re-writing of  a special case of the earlier formula using symmetries and adjoint-symmetries;
\newline
(2) Ibragimov's ``nonlinear self-adjointness'' condition in its most general form
is equivalent to the existence of an adjoint-symmetry for a general DE system,
and reduces to the existence of a symmetry in the case of a variational DE system;
\newline
(3) this formula does not always yield all admitted local conservation laws
and it produces trivial conservation laws whenever the symmetry is a translation
and the adjoint-symmetry is translation-invariant;
\newline
(4) the computation to find adjoint-symmetries (and hence to apply the formula) is just
as algorithmic as the computation of local symmetries;
\newline
(5) most importantly,
\emph{if all adjoint-symmetries are known for a given DE system (whether or not it has a variational formulation) then they can be used directly to obtain all local conservation laws, providing a kind of generalization of Noether's theorem to general DE systems.}

All of these remarks have been pointed out briefly in \Ref{Anc16},
and remark (2) has been discussed in \Ref{Zha,ZhaXie},
but it seems worthwhile to give a comprehensive discussion for all of the remarks (1)--(5),
with examples,
as the formula continues to be used in recent papers
when a complete, general method for finding all local conservation laws is available instead.
In particular, for any given DE system,
a full generalization of the content of Noether's theorem is provided by
a direct method using adjoint-symmetries,
based on the framework shown in \Ref{Olv,Mar-Alo}
and presented in an algorithmic fashion in \Ref{AncBlu97,AncBlu02a,AncBlu02b,Anc03}.
In the case when a DE system is variational,
adjoint-symmetries reduce to symmetries,
and the direct method reproduces the relationship between symmetries and conservation laws in Noether's theorem but without the need for a Lagrangian.
A detailed review and further development of this general method
appears in \Ref{Anc-review}.
Consequently,
there is no need for any kind of special methods to find local conservation laws,
just as there is no need to use special methods to find local symmetries,
because the relevant determining equations can be solved in a direct algorithmic manner.

The remainder of the present paper is organized as follows.
Remarks (1) and (2) will be demonstrated in section~\ref{sec:equiv}.
Remark (3),  along with some further consequences and properties
related to the action of symmetries,
will be explained in section~\ref{sec:symmaction}.
Remarks (4) and (5) will be briefly discussed in section~\ref{sec:method}.
Throughout,
the class of DEs $u_{tt} -u_{xx}+ a(u)(u_t^2- u_x^2) + b(u)u_t + c(u)u_x +m(u)=0$
with be used as a running example to illustrate the main points,
and the notation in Ibragimov's work will be used to allow the simplest possible comparison of the results.
Some concluding remarks are made in section~\ref{sec:remarks}.

Many examples of conservation laws of wave equations and other evolution equations
can be found in \Ref{CRC,Olv,2ndbook} and references therein.

\section{Symmetries, adjoint-symmetries, and ``nonlinear self-adjointness'' }
\label{sec:equiv}

As preliminaries, a few basic tools from variational calculus will be reviewed.
Let $x=(x^1,\ldots,x^n)$ be $n\geq 1$ independent variables,
$u=(u^1,\ldots,u^m)$ be $m\geq 1$ dependent variables,
and let $\p^k u$ denote all $k$th order partial derivatives of $u$ with respect to $x$.
Introduce an index notation for the components of $x$ and $u$:
$x^i$, $i=1,\ldots,n$; and $u^\alpha$, $\alpha=1,\ldots,m$.
In this notation,
the components of $\p^k u$ are given by $u^\alpha_{i_1\cdots i_k}$,
$\alpha=1,\ldots,m$, $i_q=1,\ldots,n$, with $q=1,\ldots,k$.
Summation is assumed over each pair of repeated indices in any expression.
The coordinate space $J=(x,u,\p u,\p^2 u,\ldots)$ is called
the \emph{jet space} associated with the variables $x,u$.
A \emph{differential function} is a locally smooth function of finitely many variables in $J$.
Total derivatives with respect to $x$ applied to differential functions are denoted
$D_i= \frac{\p}{\p x^i} + u^\alpha_{i}\frac{\p}{\p u^\alpha} + u^\alpha_{ij}\frac{\p}{\p u^\alpha_{j}} + \cdots$.

The necessary tools that will now be introduced are
the Fr\'echet derivative and its adjoint derivative,
the Euler operator and its product rule,
and the Helmholtz conditions.

Given a set of $M\geq 1$ differential functions
$f_a(x,u,\p u,\ldots,\p^N u)$, $a=1,\ldots,M$,
with differential order $N\geq 1$,
the \emph{Fr\'echet derivative} is the linearization of the functions as defined by
\begin{equation}\label{frechet}
\begin{aligned}
(\delta_w f)_a &
= \Big(\parder{}{\epsilon}f_a(x,u+\epsilon w,\p (u+\epsilon w),\ldots,\p^N(u+\epsilon w))\Big)\Big|_{\epsilon =0}
\\&
= w^\alpha\parder{f_a}{u^\alpha} + w^\alpha_{i}\parder{f_a}{u^\alpha_i} +\cdots +w^\alpha_{i_1\cdots i_N}\parder{f_a}{u^\alpha_{i_1\cdots i_N}} . 
\end{aligned}
\end{equation}
This linearization can be viewed as a local directional derivative in jet space,
corresponding to the action of a generator $\hat\X=w^\alpha\p_{u^\alpha}$
in characteristic form,
$\hat\X(f) = \delta_w f$,
where $w=(w^1(x,u,\p u,\ldots, \p^k u),\ldots,w^m(x,u,\p u,\ldots, \p^k u))$
is a set of $m$ arbitrary differential functions.

It is useful also to view the Fr\'echet derivative
as a linear differential operator acting on $w$.
Then integration by parts defines the \emph{Fr\'echet adjoint derivative}
\begin{equation}\label{adjfrechet}
(\delta^*_v f)_\alpha =
v^a\parder{f_a}{u^\alpha} -D_i\Big(v^a\parder{f_a}{u^\alpha_i}\Big) +\cdots +(-1)^N D_{i_1}\cdots D_{i_N}\Big(v^a\parder{f_a}{u^\alpha_{i_1\cdots i_N}}\Big)
\end{equation}
which is a linear differential operator acting on
a set of $M\geq 1$ arbitrary differential functions
$v = (v^1(x,u,\p u,\ldots, \p^k u),\ldots,v^M(x,u,\p u,\ldots, \p^k u))$.

These two derivatives \eqref{frechet} and \eqref{adjfrechet}
are related by
\begin{equation}\label{adjoint}
v^a(\delta_w f)_a - w^\alpha(\delta^*_v f)_\alpha = D_i \Psi^i(w,v;f)
\end{equation}
where the associated vector $\Psi^i(v,w;f)$ is given by the explicit formula
\begin{equation}\label{frechetcurrent}
\begin{aligned}
\Psi^i(w,v;f) &
= w^\alpha v^a \parder{f_a}{u^\alpha_i}
+ \big(D_j w^\alpha\big) v^a\parder{f_a}{u^\alpha_{ji}}  -w^\alpha D_j\Big(v^a\parder{f_a}{u^\alpha_{ji}}\Big) +\cdots
\\&\qquad
+\sum_{q=1}^{N} (-1)^{q-1} \big(D_{j_1}\cdots D_{j_{N-q}}w^\alpha\big) D_{i_1}\cdots D_{i_{q-1}}\Big(v^a\parder{f_a}{u^\alpha_ {j_1\cdots j_{N-q} i_1\cdots i_{q-1} i}} \Big) .
\end{aligned}
\end{equation}

The \emph{Euler operator} $E_{u^\alpha}$, or variational derivative,
is defined in terms of the Fr\'echet derivative through the variational relation
\begin{equation}\label{frechet-euler}
\delta_w f = w^\alpha E_{u^\alpha}(f) +D_i \Phi^i(w;f)
\end{equation}
which is obtained from integration by parts,
yielding
\begin{equation}\label{eulerop}
E_{u^\alpha}(f) =
\parder{f}{u^\alpha} -D_i\Big(\parder{f}{u^\alpha_i}\Big) + \cdots
+ (-1)^N D_{i_1}\cdots D_{i_N}\Big(\parder{f}{u^\alpha_{i_1\cdots i_N}}\Big)
= \dfrac{\delta f}{\delta u^\alpha}
\end{equation}
where
\begin{equation}\label{current-rel}
\Phi^i(w;f) v = \Psi^i(w,v;f) .
\end{equation}
Here, for simplicity, $f(x,u,\p u,\ldots,\p^N u)$ is a single differential function.
In particular, an explicit formula for $\Phi^i(w;f)$ is given by
\begin{equation}\label{frechet-euler-current}
\begin{aligned}
\Phi^i(w;f) & = w^\alpha\parder{f}{u^\alpha_i} + \big(D_j w^\alpha\big) \parder{f}{u^\alpha_{ji}} - w^\alpha D_j\parder{f}{u^\alpha_{ji}} +\cdots
\\&\qquad
+ \sum_{q=1}^{N}(-1)^{q-1}\big(D_{j_1}\cdots D_{j_{N-q}}w^\alpha\big) D_{i_1}\cdots D_{i_{q-1}} \parder{f}{u^\alpha_{j_1\cdots j_{N-q} i_1\cdots i_{q-1}i}}
\end{aligned}
\end{equation}
from expression \eqref{frechetcurrent}.

The Euler operator \eqref{eulerop} has the following three important properties:
First, it obeys the product rule
\begin{equation}\label{eulerop-productrule}
E_{u^\alpha}(fg) = (\delta_g^* f)_\alpha + (\delta_f^* g)_\alpha .
\end{equation}
Second, its kernel
\begin{equation}\label{eulerop-kernel}
E_{u^\alpha}(f)=0
\end{equation}
is given by total divergences
\begin{equation}\label{eulerop-div}
f=D_i F^i
\end{equation}
holding for some differential vector function $F^i$.
Third, its image consists of differential functions
\begin{equation}\label{eulerop-image}
E_{u^\alpha}(f)= g_\alpha
\end{equation}
characterized by the Helmholtz conditions
\begin{equation}\label{helmholtz}
(\delta_w g)_\alpha = (\delta^*_w g)_\alpha
\end{equation}
where $w^\alpha$ is a set of arbitrary differential functions.

There are several common alternative notations for the Fr\'echet derivative and its adjoint:
$\delta_w f =f'(w)$ and $\delta^*_v f = f'{}^*(v)$
appears in the literature on integrable systems and in \Ref{Anc-review};
$\delta_w f=D_w f$ and $\delta^*_v f=D_v^* f$
is used in Olver's book \cite{Olv};
$\delta_w f = L[u]w$
is used in the early work of Anco and Bluman \cite{AncBlu97,AncBlu02a,AncBlu02b}
and in the book \cite{2ndbook}.
In contrast, Ibragimov \cite{Ibr07,Ibr11} uses $\delta^*_v f = f^*[u,v]$.

\subsection{Conservation laws and symmetries}

Consider an $N$th-order system of $M\geq 1$ DEs
\begin{equation}\label{sys}
F=(F_1(x,u,\p u,\ldots,\p^N u),\ldots,F_M(x,u,\p u,\ldots,\p^N u))=0 .
\end{equation}
The space of solutions $u(x)$ of the system will be denoted $\Esp$.
When the number of independent variables $x$ is $n=1$,
each DE is an ordinary differential equation (ODE),
whereas when the number of independent variables $x$ is $n\geq 2$,
each DE is a partial differential equation (PDE).
The number, $m$, of dependent variables $u$ need not be the same as 
the number, $M$, of DEs in the system.

A \emph{local infinitesimal symmetry} \cite{Olv,2ndbook,1stbook}
of a given DE system \eqref{sys} is a generator
\begin{equation}\label{generator}
\X=\xi^i(x,u,\p u,\ldots,\p^r u)\parderop{x^i} +\eta^\alpha(x,u,\p u,\ldots,\p^r u)\parderop{u^\alpha}
\end{equation}
whose prolongation leaves invariant the DE system,
\begin{equation}\label{Xsymmcond}
\pr\X(F)|_\Esp =0
\end{equation}
which holds on the whole solution space $\Esp$ of the system.
(In this determining equation, 
the notation $\Esp$ means that the given DE system 
as well as its differential consequences are to be used.)
The differential functions $\xi^i$ and $\eta^\alpha$
in the symmetry generator
are called the \emph{symmetry characteristic functions}.
When acting on the solution space $\Esp$,
an infinitesimal symmetry generator can be formally exponentiated to produce
a one-parameter group of transformations $\exp(\epsilon\pr\X)$,
with parameter $\epsilon$,
where the infinitesimal transformation is given by
\begin{equation}\label{symmgroup-action}
\begin{aligned}
u^\alpha(x) \rightarrow u^\alpha(x)
 + & \epsilon\big( \eta^\alpha(x,u(x),\p u(x),\ldots,\p^r u(x))
-u^\alpha_i(x)\xi^i(x,u(x),\p u(x),\ldots,\p^r u(x)) \big)
\\&\qquad
+O\big(\epsilon^2\big)
\end{aligned}
\end{equation}
for all solutions $u(x)$ of the DE system.

Two infinitesimal symmetries are equivalent if they have the same action \eqref{symmgroup-action}
on the solution space $\Esp$ of a given DE system.
An infinitesimal symmetry is thereby called \emph{trivial}
if it leaves all solutions $u(x)$ unchanged.
This occurs iff its characteristic functions satisfy the relation
\begin{equation}
\eta^\alpha|_\Esp=(u^\alpha_{i}\xi^i)|_\Esp .
\end{equation}
The corresponding generator \eqref{generator} of a trivial symmetry
is thus given by
\begin{equation}\label{trivsymm}
\X_\triv|_\Esp= \xi^i \parderop{x^i} + \xi^i u^\alpha_{i}\parderop{u^\alpha}
\end{equation}
which has the prolongation $\pr\X_\triv|_\Esp=\xi^i D_i$.
Conversely, any generator of this form on the solution space $\Esp$
represents a trivial symmetry.
Thus, any two generators that differ by a trivial symmetry are equivalent. 
The \emph{differential order of an infinitesimal symmetry} is defined to be
the smallest differential order among all equivalent generators.

Any symmetry generator is equivalent to a generator given by
\begin{equation}\label{symmchar}
\hat\X=\X-\X_\triv = P^\alpha\parderop{u^\alpha},
\quad
P^\alpha =\eta^\alpha -\xi^i u^\alpha_i, 
\end{equation}
under which $u$ is infinitesimally transformed while $x$ is invariant,
due to the relation
\begin{equation}\label{XhatXrel}
\pr\X-\pr\hat\X= \xi^i D_i .
\end{equation}
This generator \eqref{symmchar} defines the \emph{characteristic form}
for the infinitesimal symmetry.
The symmetry invariance \eqref{Xsymmcond} of the DE system
can then be expressed by
\begin{equation}\label{Xsymmdeteq}
\pr\hat\X(F)|_\Esp = 0
\end{equation}
holding on the whole solution space $\Esp$ of the given system.
Note that the action of $\pr\hat\X$ is the same as a Fr\'echet derivative \eqref{frechet},
and hence an equivalent, modern formulation \cite{Olv,2ndbook,Anc-review} of this invariance \eqref{Xsymmdeteq}
is given by the \emph{symmetry determining equation}
\begin{equation}\label{symmdeteq}
(\delta_P F)_a|_\Esp = 0 .
\end{equation}
(Recall, the notation $\Esp$ means that the given DE system 
as well as its differential consequences are to be used in these determining equations.)

In jet space $J$,
a group of transformations $\exp(\epsilon\pr\X)$ with a non-trivial generator $\X$
in general will not act in a closed form
on $x,u$, and derivatives $\p^k u$ up to a finite order,
except \cite{Olv,2ndbook} for point transformations acting on $(x,u)$,
and contact transformations acting on $(x,u,\p u)$.
Moreover, a contact transformation is a prolonged point transformation
when the number of dependent variables is $m>1$ \cite{Olv,2ndbook}.
A \emph{point symmetry} is defined as a symmetry transformation group on $(x,u)$,
whose generator is given by characteristic functions of the form
\begin{equation}\label{Xpointsymm}
\X=\xi(x,u)^i\parderop{x^i} +\eta^\alpha(x,u)\parderop{u^\alpha}
\end{equation}
corresponding to the infinitesimal point transformation
\begin{equation}\label{pointgroup}
x^i\rightarrow x^i+\epsilon\, \xi^i(x,u) + O(\epsilon^2),
\quad
u^\alpha\rightarrow u^\alpha+\epsilon\, \eta^\alpha(x,u) + O(\epsilon^2) .
\end{equation}
Likewise,
a \emph{contact symmetry} is defined as a symmetry transformation group on $(x,u,\p u)$
whose generator corresponds to an infinitesimal transformation that preserves
the contact relations $u^\alpha_i =D_i u^\alpha$.
The set of all admitted point symmetries and contact symmetries
for a given DE system comprises its group of \emph{Lie symmetries}.
The corresponding generators of this group comprise a Lie algebra \cite{Olv,1stbook,2ndbook}. 

A \emph{local conservation law} of a given DE system \eqref{sys}
is a divergence equation
\begin{equation}\label{conslaw}
D_i C^i|_\Esp =0
\end{equation}
which holds on the whole solution space $\Esp$ of the system,
where
\begin{equation}\label{conscurrent}
C=(C^1(x,u,\p u,\ldots,\p^r u),\ldots,C^n(x,u,\p u,\ldots,\p^r u))
\end{equation}
is the \emph{conserved current vector}.
In the case when one of the independent variables represents a time coordinate,
and the remaining $n-1$ independent variables represent space coordinates,
namely $x=(t,x^1,\ldots,x^{n-1})$,
then $C^1=T$ is a \emph{conserved density}
and $(C^2,\ldots,C^n)=\vec X$ is a \emph{spatial flux vector},
while the conservation law has the form of a local continuity equation
$(D_t T + \Div \vec X)|_\Esp =0$.
(Similarly to the symmetry determining equation, 
the notation $\Esp$ here means that the given DE system 
as well as its differential consequences are to be used.)

A conservation law \eqref{conslaw} is \emph{locally trivial} if
\begin{equation}\label{trivconslaw}
C^i|_\Esp = D_j\Theta^{ij}
\end{equation}
holds for some differential antisymmetric tensor function
$\Theta^{ij}(x,u,\p u,\ldots,\p^{r-1} u)$ on $\Esp$, 
since any total curl is identically divergence free,
$D_i(D_j\Theta^{ij}) = D_iD_j\Theta^{ij} = 0$ due to commutativity of total derivatives.
Two conservation laws are said to be \emph{locally equivalent}
if, on the solution space $\Esp$,
their conserved currents differ by a locally trivial current \eqref{trivconslaw}.
The \emph{differential order of a conservation law} is defined to be
the smallest differential order among all locally equivalent conserved currents.
(Sometimes a local conservation law is itself defined
as the equivalence class of locally equivalent conserved currents.)

For a given DE system \eqref{sys},
the set of all non-trivial local conservation laws (up to local equivalence)
forms a vector space on which the local symmetries of the system have a natural action
\cite{Kha,Olv,2ndbook}.
In particular,
the infinitesimal action of a symmetry \eqref{generator}
on a conserved current \eqref{conscurrent} is given by \cite{Olv}
\begin{equation}\label{X_on_C}
C^i_\X = \pr\X(C^i) + C^i D_i\xi^i -C^jD_j\xi^i .
\end{equation}
When the symmetry is expressed in characteristic form \eqref{symmchar},
its action has the simple form
\begin{equation}\label{symm_on_C}
C^i_{\hat\X} = \pr\hat\X(C^i)=\delta_P C^i .
\end{equation}
The conserved currents $C^i_\X$ and $C^i_{\hat\X}$ are locally equivalent,
\begin{equation}
(C^i_{\hat\X} - C^i_\X)|_\Esp = D_j\Theta^{ij}
\end{equation}
with
\begin{equation}
\Theta^{ij} = \xi^i C^j-\xi^j C^i
\end{equation}
which follows from the relation \eqref{XhatXrel}.

A DE system is \emph{variational} if it arises as the Euler-Lagrange equations of
a local Lagrangian.
This requires that the number of equations in the system is the same as the number of dependent variables, $M=m$,
and that the differential order $N$ of the system is even,
in which case the system is given by
\begin{equation}\label{variational_sys}
F_\alpha = E_{u^\alpha}(L),
\quad
\alpha =1,\ldots,M=m
\end{equation}
where the Lagrangian is a differential function
\begin{equation}\label{variational_sys_L}
L(x,u,\p u,\ldots, \p^{N/2} u) .
\end{equation}
The necessary and sufficient conditions \cite{Olv,1stbook,2ndbook,Anc-review}
for a given DE system \eqref{sys} to be variational 
consist of the Helmholtz conditions \eqref{helmholtz},
which are given by
\begin{equation}\label{helmholtz_conds}
(\delta_w F)_\alpha = (\delta^*_w F)_\alpha
\end{equation}
where $w^\alpha$ is a set of arbitrary differential functions.
Note these conditions \eqref{helmholtz_conds} are required to hold identically in jet space $J$ 
(and not just on the solution space $\Esp$ of the DE system).

\subsection{Ibragimov's conservation law formula}

The starting point is the well-known observation \cite{Olv} that
any $N$th-order system of $M\geq 1$ DEs \eqref{sys}
can be embedded into a larger system
by appending an ``adjoint variable'' for each DE in the system,
where this set of $M\geq 1$ variables $v=(v^1,\ldots,v^M)$
is taken to satisfy the adjoint of the linearization of the original DE system.
Specifically, the enlarged DE system is given by
\begin{align}
& F_a(x,u,\p u,\ldots,\p^N u) =0,
\quad
a=1,\ldots,M
\label{u_sys}
\\
&
(\delta^*_v F)_\alpha =F^*_\alpha(x,u,v,\p u,\p v,\ldots,\p^N u,\p^N v) =0,
\quad
\alpha=1,\ldots,m
\label{v_sys}
\end{align}
for $u^\alpha(x)$ and $v^a(x)$,
in Ibragimov's notation.
This system \eqref{u_sys}--\eqref{v_sys} comprises
the Euler-Lagrange equations of the Lagrangian function
\begin{equation}\label{L}
L = v^a F_a(x,u,\p u,\ldots,\p^N u)
\end{equation}
since, clearly,
\begin{equation}\label{EL_sys}
E_{v^a}(L)= F_a,
\quad
E_{u^\alpha}(L)= (\delta^*_v F)_\alpha
\end{equation}
through the product rule \eqref{eulerop-productrule}.

All solutions $u(x)$ of the original DE system \eqref{u_sys}
give rise to solutions of the Euler-Lagrange system \eqref{EL_sys}
by letting $v(x)$ be any solution (for instance $v=0$) of the DEs \eqref{v_sys}.
Conversely,
all solutions $(u(x),v(x))$ of the Euler-Lagrange system \eqref{EL_sys}
yield solutions of the original DE system \eqref{u_sys} by projecting out $v(x)$.

This embedding relationship can used to show that
every symmetry of the original DE system \eqref{u_sys}
can be extended to a variational symmetry of the Euler-Lagrange system \eqref{EL_sys}.
The proof is simplest when the symmetries are formulated in characteristic form \eqref{symmchar}.

Let
\begin{equation}\label{symm}
\hat\X=P^\alpha(x,u,\p u,\ldots,\p^r u)\parderop{u^\alpha}
\end{equation}
be any local symmetry generator (in characteristic form)
admitted by the DE system \eqref{u_sys}.
Under some mild regularity conditions \cite{Anc-review} on the form of these DEs,
the symmetry determining equation \eqref{symmdeteq} implies that
the characteristic functions $P^\alpha$ satisfy
\begin{equation}\label{symmdeteq_offE}
(\delta_P F)_a = R_P(F)_a
\end{equation}
where
\begin{equation}\label{symm_op}
R_P= R_P{}^{b}_{a} + R_P{}^{b i}_{a} D_{i} + R_P{}^{b ij}_{a} D_{i} D_{j} + \cdots + R_P{}^{b i_1\cdots i_r}_{a} D_{i_1}\cdots D_{i_r}
\end{equation}
is some linear differential operator whose coefficients
$R_P{}^{b}_{a},R_P{}^{b i}_{a},\ldots,R_P{}^{b i_1\cdots i_r}_{a}$
are differential functions that are non-singular
on solution space $\Esp$ of the DE system \eqref{sys}.
Now consider the action of this symmetry generator \eqref{symm}
on the Lagrangian \eqref{L}.
From the operator relation \eqref{symmdeteq_offE} followed by integration by parts,
the symmetry action is given by
\begin{equation}\label{hatX_on_L}
\pr\hat\X(L) = v^a R_P(F)_a = F_a R_P^*(v)^a + D_i\hat\Theta^i
\end{equation}
where
\begin{equation}\label{symm_adjop}
R_P^*= R_P^*{}^{b}_{a} -R_P^*{}^{b i}_{a} D_{i} +R_P^*{}^{b ij}_{a} D_{i} D_{j}+ \cdots +(-1)^r R_P^*{}^{b i_1\cdots i_r}_{a} D_{i_1}\cdots D_{i_r}
\end{equation}
is the adjoint of the operator \eqref{symm_op},
with the non-singular coefficients
\begin{equation}
\begin{aligned}
& R_P^*{}^{b}_{a}  = R_P{}^{b}_{a}  -(D_j R_P{}^{bj}_{a}) +\cdots +(-1)^r (D_{j_1}\cdots D_{j_r} R_P{}^{b j_1\cdots j_r}_{a})  ,
\\
& R_P^* {}^{bi}_{a}= R_P{}^{bi}_{a}  -\tbinom{2}{1}(D_jR_P{}^{bji}_{a}) +\cdots + (-1)^{r-1}\tbinom{r}{r-1} (D_{j_1}\cdots D_{j_{r-1}}R_P{}^{bj_1\cdots j_{r-1}i}_{a}) ,
\\
& R_P^* {}^{bij}_{a}= R_P{}^{bij}_{a}  -\tbinom{3}{1}(D_kR_P{}^{bkij}_{a}) +\cdots + (-1)^{r-2}\tbinom{r}{r-2} (D_{j_1}\cdots D_{j_{r-2}}R_P{}^{bj_1\cdots j_{r-2}ij}_{a}) ,
\\
& \quad\vdots
\\
& R_P^*{}^{bi_1\cdots i_r}_{a} = R_P{}^{bi_1\cdots i_r}_{a} D_{i_1}\cdots D_{i_r} .
\end{aligned}
\end{equation}
Although the Lagrangian is not preserved,
the expression \eqref{hatX_on_L} for the symmetry action shows that
if the symmetry is extended to act on $v$ via
\begin{equation}\label{hatXext}
\hat\X^\ext=P^\alpha \parderop{u^\alpha} - R_P^*(v)^a\parderop{v^a} ,
\end{equation}
then under this extended symmetry the Lagrangian will be invariant up to a total divergence,
\begin{equation}\label{hatXext_on_L}
\pr\hat\X^\ext(L) = v^a R_P(F)_a - F_a R_P^*(v)^a = D_i\hat\Theta^i .
\end{equation}
This completes the proof.
A useful remark is that the vector $\hat\Theta^i$ in the total divergence \eqref{hatXext_on_L}
is a linear expression in terms of $F_a$ (and total derivatives of $F_a$),
and hence this vector vanishes
whenever $u(x)$ is a solution of the DE system \eqref{u_sys}.
Consequently, $\hat\Theta^i$ is a trivial current for the Euler-Lagrange system \eqref{EL_sys}.

Some minor remarks are that the proof given by Ibragimov \cite{Ibr07}
does not take advantage of the simplicity of working with symmetries in characteristic form
and also glosses over the need for some regularity conditions on the DE system
so that the symmetry operator relation \eqref{symmdeteq_offE} will hold.
Moreover, that proof is stated only for DE systems in which
the number of equations is the same as the number of dependent variables, $M=m$.

Now, since the extended symmetry \eqref{hatXext} is variational,
Noether's theorem can be applied to obtain a corresponding conservation law
for the Euler-Lagrange system \eqref{EL_sys},
without the need for any additional conditions.
The formula in Noether's theorem
comes from applying the variational identity \eqref{frechet-euler} to the Lagrangian \eqref{L},
which yields
\begin{equation}\label{var_identity}
\pr\hat\X(L) = \phi^aF_a + v^a(\delta_P F)_a
= \hat\X(v^a) E_{v^a}(L) + \hat\X(u^\alpha)E_{u^\alpha}(L) + D_i\Phi^i(P;L)
\end{equation}
for any generator
\begin{equation}\label{hatX_uv}
\hat\X = P^\alpha\parderop{u^\alpha} + \phi^a\parderop{v^a} .
\end{equation}
The total divergence term $D_i\Phi^i(P;L)$
is given by the formula \eqref{frechet-euler-current} derived using the Euler operator
\eqref{eulerop}.
This yields
\begin{equation}\label{L-current}
\begin{aligned}
\Phi^i(P;L)
& = P^\alpha v^a \parder{F_a}{u^\alpha_i}
+ \big(D_j P^\alpha\big) v^a\parder{F_a}{u^\alpha_{ji}}  -P^\alpha D_j\Big(v^a\parder{F_a}{u^\alpha_{ji}}\Big) +\cdots
\\&\qquad
+\sum_{q=1}^{N} (-1)^{q-1} \big(D_{j_1}\cdots D_{j_{N-q}}P^\alpha\big) D_{i_1}\cdots D_{i_{q-1}}\Big(v^a\parder{F_a}{u^\alpha_ {j_1\cdots j_{N-q} i_1\cdots i_{q-1} i}} \Big) .
\end{aligned}
\end{equation}
When this variational identity \eqref{var_identity} is combined with
the action \eqref{hatXext_on_L} of the variational symmetry \eqref{hatXext} on the Lagrangian,
the following Noether relation is obtained:
\begin{equation}\label{noether_rel}
D_i(\hat\Psi^i-\Phi^i(P;L) = \phi^a F_a + P^\alpha F^*_\alpha,
\quad
\phi^a = - R_P^*(v)^a
\end{equation}
where $F^*_\alpha$ is expression \eqref{v_sys}.
Since $F_a$, $F^*_\alpha$, and $\hat\Psi^i$ vanish
when $(u(x),v(x))$ is any solution of the Euler-Lagrange system \eqref{EL_sys},
the Noether relation \eqref{noether_rel} yields a local conservation law
\begin{equation}\label{uv_conslaw}
D_i \hat C^i|_{\Esp(u,v)}=0,
\quad
\hat C^i=\Phi^i(P;L)
\end{equation}
where $\Esp(u,v)$ denotes the solution space of the system \eqref{EL_sys}
(including its differential consequences). 
This conservation law is locally equivalent to the conservation law formula
underlying Ibragimov's work \cite{Ibr07,Ibr11},
which is given by
\begin{equation}\label{uv_conslaw_equiv}
D_i C^i|_{\Esp(u,v)}=0,
\quad
C^i=\hat C^i -\xi^i L
\end{equation}
where $C^i|_{\Esp(u)}=\hat C^i|_
{\Esp(u)}$ since $L|_{\Esp(u)}=0$.
Strangely,
nowhere does Ibragimov (or subsequent authors)
point out that the term $\xi^i L$ in the conserved current
trivially vanishes on all solutions $(u(x),v(x))$ of the Euler-Lagrange system!

Hence, the following result has been established.

\begin{prop}\label{noetherformula}
Any DE system \eqref{u_sys} can be embedded into a larger Euler-Lagrange system \eqref{EL_sys}
such that every symmetry \eqref{symm} of the original system
can be extended to a variational symmetry \eqref{hatXext} of the Euler-Lagrange system.
Noether's theorem then yields a conservation law \eqref{uv_conslaw}
for all solutions $(u(x),v(x))$ of the Euler-Lagrange system \eqref{EL_sys}.
\end{prop}

A side remark is that the locally equivalent conservation law \eqref{uv_conslaw_equiv}
also can be derived from Noether's theorem
if the extended symmetry \eqref{hatXext} is expressed in canonical form
\begin{equation}\label{Xext}
\X^\ext=\xi^i\parderop{x^i} + \eta^\alpha \parderop{u^\alpha} +(\xi^iv^a_i - R_P^*(v)^a)\parderop{v^a}
\end{equation}
as obtained from relations \eqref{symmchar}--\eqref{XhatXrel}.
In particular,
the corresponding form of the variational identity \eqref{var_identity}
becomes
\begin{equation}\label{Xext_var_id}
\pr\X^\ext(L) +(D_i\xi^i)L
= \hat\X(v^a) E_{v^a}(L) + \hat\X(u^\alpha)E_{u^\alpha}(L) + D_i(\xi^iL+\Phi^i(P;L))
\end{equation}
where $P^\alpha =\eta^\alpha -\xi^i u^\alpha_i$ and $Q^a= - R_P^*(v)^a$,
while the action of the symmetry \eqref{Xext} on the Lagrangian is given by
\begin{equation}\label{Xext_on_L_equiv}
\begin{aligned}
\pr\X^\ext(L) +(D_i\xi^i)L
& = (D_i(\xi^i v^a) - R_P^*(v)^a) F_a + v^a \pr\X^\ext(F^a)
\\
& = (D_i(\xi^i v^a) - R_P^*(v)^a) F_a + v^a (R_P(F)^a + \xi^iD_i F^a)  =D_i\Theta^i
\end{aligned}
\end{equation}
where $\Theta^i$ vanishes whenever $u(x)$ is a solution of the DE system \eqref{u_sys}.
Hence,
the Noether relation obtained from
combining equations \eqref{Xext_var_id}--\eqref{Xext_on_L_equiv}
yields the conserved current $C^i =\Phi^i(P,v;F) -\xi^iL$
modulo the locally trivial current $\Theta^i$.
If the original symmetry \eqref{symm} being used is a point symmetry,
then this trivial current $\Theta^i$ can be shown to vanish identically,
which is the situation considered in  Ibragimov's papers \cite{Ibr07,Ibr11}
and in nearly all subsequent applications in the literature.

\subsection{``Nonlinear self-adjointness''}

The conservation law \eqref{uv_conslaw} holds for all solutions $(u(x),v(x))$
of the Euler-Lagrange system \eqref{EL_sys}.
It seems natural to restrict this
to solutions of the original DE system \eqref{u_sys} for $u(x)$ by putting $v=0$.
However, the resulting conserved current is trivial,
$\Phi^i(P;L)|_{v=0}= \Phi^i(P;0)=0$,
because $L$ is a linear expression in terms of $v$.
Consequently,
some other way must be sought to project the solution space $\Esp(u,v)$ of the Euler-Lagrange system
onto the solution space $\Esp$ of the original DE system \eqref{u_sys}.

Ibragimov's first paper \cite{Ibr07} proposes to put $v=u$,
which is clearly a significant restriction on the form of the original DE system \eqref{u_sys}.
In particular,
this requires that $F^*_\alpha|_{v=u}=F_a$ hold identically,
where the DE system is assumed to have the same number of equations
as the number of dependent variables, $M=m$,
which allows the indices $a=\alpha$ to be identified.
He calls such a DE system $F_a=0$ ``strictly self-adjoint''.
This definition is motivated by the case of a linear DE system,
since linearity implies
$(\delta_u F)_a = F_a$ and $(\delta^*_u F)_\alpha =F^*_\alpha|_{v=u}$
are identities,
whereby a linear DE system with $M=m$ is ``strictly self-adjoint'' iff
it satisfies $(\delta F)_\alpha = (\delta^* F)_\alpha$
which is the condition for self-adjointness of a linear system.
However, for nonlinear DE systems,
the definition of ``strictly self-adjoint''
conflicts with the standard of definition \cite{Olv,KraVin} in variational calculus that
a general DE system $F_a=0$ is self-adjoint iff
its associated Fr\'echet  derivative operator is self-adjoint,
$(\delta F)_a = (\delta^* F)_\alpha$,
which requires $M=m$.

Ibragimov subsequently \cite{Ibr07b} proposed to have $v=\phi(u)$,
which he called ``quasi-self-adjointness''. 
A more general proposal $v=\phi(x,u)$ was then introduced first in \Ref{Gan11}
and shortly later appears in Ibragimov's next paper \cite{Ibr11},
with the condition that $F^*_\alpha|_{v=\phi(x,u)}=\lambda_\alpha{}^\beta F_\beta$
must hold for some coefficients $\lambda_\alpha{}^\beta$,
again with $M=m$.
This condition is called ``weak self-adjointness'' in \Ref{Gan11}
and ``nonlinear self-adjointness'' in \Ref{Ibr11}.
Ibragimov also mentions an extension of this definition to
$v=\phi(x,u,\p u, \ldots, \p^s u)$, but does not pursue it. 
Later he applies this definition in \Ref{GalIbr} to a specific PDE, 
where $\lambda_\alpha{}^\beta$ is extended to be a linear differential operator.
However, unlike in the previous papers, 
no conservation laws are found from using this extension. 
A subsequent paper \cite{Gan14} then uses this extension,
which is called ``nonlinear self-adjointness through a differential substitution'', 
to obtain conservation laws for several similar PDEs. 
Finally, the same definition is stated more generally in \Ref{ZhaXie} 
for DE systems with $M=m$:
\begin{equation}
F^*_\alpha|_{v=\phi(x,u,\p u,\ldots,\p^s u)}=
\lambda_\alpha{}^\beta F_\beta + \lambda_\alpha{}^{\beta i}D_i F_\beta  + \cdots
+ \lambda_\alpha{}^{\beta i_1\cdots i_r}D_{i_1}\cdots D_{i_p} F_\beta
\end{equation}
where the coefficients
$\lambda_\alpha{}^\beta,\lambda_\alpha{}^{\beta i},\ldots,\lambda_\alpha{}^{\beta i_1\cdots i_r}$
are differential functions.

These developments lead to the following conservation law theorem,
which is a generalization of Ibragimov's main theorem \cite{Ibr07,Ibr11}
to arbitrary DE systems (not restricted by $M=m$),
combined with the use of a differential substitution \cite{Ibr11,Gan14,ZhaXie}.

\begin{thm}\label{Ibr-thm}
Suppose a system of DEs \eqref{sys} satisfies
\begin{equation}\label{nonlin-self-adj}
F^*_\alpha|_{v=\phi}=
\lambda_\alpha{}^a F_a + \lambda_\alpha{}^{a i}D_i F_a +\cdots + \lambda_\alpha{}^{a i_1\cdots i_pr}D_{i_1}\cdots D_{i_p} F_a
\end{equation}
for some differential functions
$\phi^a(x,u,\p u,\ldots,\p^s u)$
and
$\lambda_\alpha{}^a(x,u,\p u,\ldots,\p^s u)$,
$\lambda_\alpha{}^{a i}(x,u,\p u,\ldots,\p^s u)$,
$\ldots$,
$\lambda_\alpha{}^{a i_1\cdots i_p}(x,u,\p u,\ldots,\p^s u)$
that are non-singular on the solution space $\Esp$ of the DE system,
where $F^*_\alpha$ is the adjoint linearization \eqref{adjfrechet} of the system.
Then any local symmetry
\begin{equation}\label{hatX_u}
\X = \xi^i(x,u,\p u,\ldots,\p^r u)\parderop{x^i} + \eta^\alpha(x,u,\p u,\ldots,\p^r u)\parderop{u^\alpha}
\end{equation}
admitted by the DE system
yields a local conservation law \eqref{conslaw} given in an explicit form by
the conserved current \eqref{L-current}
with $v^a= \phi^a$ and $P^\alpha=\eta^\alpha-\xi^i u^\alpha_i$.
\end{thm}

An important remark is that all of the functions
$\phi^a$, $\lambda_\alpha{}^a,\lambda_\alpha{}^{a i},\ldots,\lambda_\alpha{}^{a i_1\cdots i_p}$
must be non-singular on $\Esp$,
as otherwise the condition \eqref{nonlin-self-adj} can be satisfied in a trivial way.
This point is not mentioned in any of the previous work
\cite{Ibr07,Ibr11,Gan11,Gan14,ZhaXie}.

The ``nonlinear self-adjointness'' condition \eqref{nonlin-self-adj}
turns out to have a simple connection to the determining equations for symmetries.
This connection is somewhat obscured by the unfortunate use of
non-standard definitions and non-standard notation in \Ref{Ibr07,Ibr11}.
Nevertheless, 
it is straightforward to show that equation \eqref{nonlin-self-adj} is precisely
the adjoint of the determining equation \eqref{symmdeteq} for symmetries
formulated as an operator equation \eqref{symmdeteq_offE}.

\subsection{Adjoint-symmetries and a formula for generating conservation laws}

For any given DE system \eqref{sys},
the adjoint of the symmetry determining equation \eqref{symmdeteq} is given by
\begin{equation}\label{adjsymmdeteq}
(\delta^*_Q F)_\alpha|_\Esp = 0
\end{equation}
for a set of differential functions $Q^a(x,u,\p u,\ldots,\p^r u)$.
(Similarly to the symmetry determining equation, 
the notation $\Esp$ here means that the given DE system 
as well as its differential consequences are to be used.)
These differential functions are called an \emph{adjoint-symmetry} \cite{AncBlu97},
in analogy to the characteristic functions of a symmetry \eqref{symm},
and so the equation \eqref {adjsymmdeteq}
is called the \emph{adjoint-symmetry determining equation}.
As shown in \Ref{Anc-review},
this analogy has a concrete geometrical meaning
in the case when a DE system is an evolutionary system
$F_\alpha = u^\alpha_t - f_\alpha(x,u,\p_x u,\ldots,\p_x^N u) =0$
with $M=m$ and  $x=(t,x^1,\ldots,x^{n-1})$,
where $t$ is a time coordinate and $x^i$, $i=1,\ldots,n-1$, are space coordinates.
In this case,
$Q^\alpha$ can be viewed as the coefficients of a $1$-form or a covector
$Q^\alpha \extder{u^\alpha}$,
in analogy to $P^\alpha$ being the coefficients of
a vector $P^\alpha\parderop{u^\alpha}$.
The condition for $P^\alpha\parderop{u^\alpha}$ to be a symmetry can be formulated as
$\big(\lieder{f} P^\alpha\parderop{u^\alpha}\big)|_\Esp=0$
where $\lieder{f}$ denotes the Lie derivative \cite{Olv,Anc-review}
with respect to the time evolution vector $\Y= f_\alpha \parderop{u^\alpha}$.
Then the condition for $Q^\alpha \extder{u^\alpha}$ to be an adjoint-symmetry
is equivalent to $\big(\lieder{f} Q^\alpha\extder{u^\alpha}\big)|_\Esp=0$.
(Note the awkwardness in the index positions here comes from
Ibragimov's choice of index placement $F_\alpha$ for a DE system with $M=m$.
A better notation would be $F^\alpha$, and $F^a$ when $M\neq m$,
which is used in \Ref{AncBlu97,AncBlu02a,AncBlu02b,2ndbook}.)

In the case when a DE system is variational \eqref{variational_sys},
the symmetry determining equation is self-adjoint,
since $(\delta^*_Q F)_\alpha = (\delta_Q F)_\alpha$.
Then the adjoint-symmetry determining equation \eqref{adjsymmdeteq}
reduces to the symmetry determining equation \eqref{symmdeteq},
with $Q^a(x,u,\p u,\ldots,\p^r u) = P^\alpha(x,u,\p u,\ldots,\p^r u)$,
where the indices $a=\alpha$ can be identified, due to $M=m$.
Consequently,
adjoint-symmetries of any variational DE system are the same as symmetries.

Other aspects of adjoint-symmetries and their connection to symmetries are discussed
in \Ref{Anc17}.

Now, under some mild regularity conditions \cite{Anc-review}
on the form of a general DE system \eqref{sys},
the adjoint-symmetry determining equation \eqref{adjsymmdeteq}
implies that the functions $Q^a$ satisfy
\begin{equation}\label{adjsymmdeteq_offE}
(\delta^*_Q F)_\alpha = R_Q(F)_\alpha
\end{equation}
where
\begin{equation}\label{adjsymm_op}
R_Q= R_Q{}^{b}_{\alpha} + R_Q{}^{b i}_{\alpha} D_{i} + R_Q{}^{b ij}_{\alpha} D_{i} D_{j} + \cdots + R_Q{}^{b i_1\cdots i_r}_{\alpha} D_{i_1}\cdots D_{i_r}
\end{equation}
is some linear differential operator whose coefficients
$R_Q{}^{b}_{\alpha},R_Q{}^{b i}_{\alpha},\ldots,R_Q{}^{b i_1\cdots i_r}_{\alpha}$
are differential functions that are non-singular
on solution space $\Esp$ of the DE system \eqref{sys}.
In Ibragimov's notation
$F^*_\alpha(x,u,v,\p u,\p v,\ldots,\p^N u,\p^N v) = (\delta^*_v F)_\alpha$,
the adjoint-symmetry equation \eqref{adjsymmdeteq_offE}
coincides with the ``nonlinear self-adjointness'' condition \eqref{nonlin-self-adj}
in Theorem~\ref{Ibr-thm},
where the operator on the right-hand side of equation \eqref{nonlin-self-adj}
is precisely the adjoint-symmetry operator \eqref{adjsymm_op}.

Therefore, the following equivalence has been established.

\begin{prop}\label{equiv-nonlin-self-adj}
For a general DE system \eqref{sys},
the condition \eqref{nonlin-self-adj} of ``nonlinear self-adjointness''
coincides with the condition of existence of an adjoint-symmetry \eqref{adjsymmdeteq}.
When a DE system is variational \eqref{variational_sys},
these conditions reduce to the condition of existence of a symmetry.
\end{prop}

One remark is that the formulation of ``nonlinear self-adjointness''  given here
is more general than what appears in \Ref{Ibr11,Gan14,ZhaXie}
since those formulations assume that the DE system has
the same number of equations as the number of dependent variables, $M=m$.
Another remark is that the meaning of ``nonlinear self-adjointness''
shown here in the case of variational DE systems has not previously
appeared in the literature.

{\bf Example}:
Consider the class of semilinear wave equations
$u_{tt} -u_{xx}+ a(u)(u_t^2- u_x^2) + b(u)u_t + c(u)u_x +m(u)=0$
for $u(t,x)$, with a nonlinearity coefficient $a(u)$, damping coefficients $b(u),c(u)$,
and a mass-type coefficient $m(u)$.
In \Ref{IbrTorTra10},
conditions under which a slightly more general family of wave equations is
``nonlinearly self-adjoint'' \eqref{nonlin-self-adj} are stated for $v=\phi(u)$.
These results will be generalized here by considering $v=\phi(t,x,u)$.
A first observation is that this class of wave equations
admits an equivalence transformation $u\to \tilde u= f(u)$, with $f'\neq0$,
which can be used to put $a=0$ by $f(u)= \int\exp(A(u))du$
where $A'=a$.
(Equivalence transformations were not considered in \Ref{IbrTorTra10},
and so their results are considerably more complicated than is necessary.)
This transformation gives
\begin{equation}\label{ex-eqn}
u_{tt} -u_{xx}+ b(u)u_t + c(u)u_x +m(u)=0 .
\end{equation}
In Ibragimov's notation,
the condition of ``nonlinear self-adjointness''  with $v=\phi(t,x,u)$
is given by
$0=E_u(vF)|_{v=\phi}$
where
\begin{equation}\label{ex-F}
F=u_{tt} -u_{xx}+ b(u)u_t + c(u)u_x +m(u) .
\end{equation}
This yields
\begin{equation}\label{ex-nonlinselfadj}
\big( D_t^2\phi  -D_x^2\phi -bD_t\phi -cD_x\phi +m'\phi \big)\big|_{F=0}=0 .
\end{equation}
For comparison,
the determining equation \eqref{symmdeteq} for local symmetries
$\hat\X=P(t,x,u,u_t,u_x,\ldots)\parderop{u}$ (in characteristic form)
is given by
\begin{equation}\label{ex-symmeq}
\big( D_t^2 P  -D_x^2 P + b D_t P  +c D_x P + (u_tb' + u_xc' +m')P =0 \big)\big|_{F=0}=0 .
\end{equation}
Its adjoint is obtained by multiplying by $Q(t,x,u,u_t,u_x,\ldots)$ and integrating by parts,
which yields
$\big( D_t^2 Q  -D_x^2 Q -D_t(b Q) -D_x(c Q) + (u_tb' + u_xc' +m')Q =0 \big)\big|_{F=0}=0$.
After the $D_x$ terms are expanded out,
this gives the determining equation \eqref{adjsymmdeteq} for local adjoint-symmetries
\begin{equation}\label{ex-adjsymmeq}
\big( D_t^2Q  -D_x^2Q -b D_tQ -c D_x Q +(b' u_t+c' u_x+m')Q \big)\big|_{F=0}=0
\end{equation}
which coincides with the ``nonlinear self-adjointness'' condition \eqref{ex-nonlinselfadj}
extended to differential substitutions \cite{Gan11,Ibr11,Zha}
given by $v=Q(t,x,u,u_t,u_x,\ldots)$.
All adjoint-symmetries of lowest-order form $Q(t,x,u)$
can be found in a straightforward way.
After $Q(t,x,u)$ is substituted into the determining equation \eqref{ex-adjsymmeq},
and $u_{tt}$ is eliminated through the wave equation \eqref{ex-eqn},
the determining equation splits with respect to the variables $u_t$ and $u_x$,
yielding a linear overdetermined system of four equations (after some simplifications):
\begin{subequations}\label{ex-adjsymm-sys}
\begin{align}
& Q_{tt} -Q_{xx} -bQ_t -cQ_x -m Q_u +m' Q=0,
\\
& Q_{tu} -bQ_u =0,
\quad
Q_{xu} +cQ_u =0,
\\
& Q_{uu} =0 .
\end{align}
\end{subequations}
It is straightforward to derive and solve this determining system by Maple.
Hereafter, the conditions
\begin{equation}\label{ex-conds}
b'\neq0,
\quad
c'\neq0,
\quad
m''\neq0,
\quad
m(0)=0
\end{equation}
will be imposed,
which corresponds to studying wave equations \eqref{ex-eqn}
whose lower-order terms are nonlinear and homogeneous.
The general solution of the determining system \eqref{ex-adjsymm-sys}
then comprises three distinct cases (as obtained using the Maple package 'rifsimp'),
after merging.
This leads to the following complete classification of solution cases
shown in table~\ref{ex-table:adjsymm}.
The table is organized by listing each solution $Q$ and the conditions on $b,c,m$ for which it exists. 
(From these conditions, 
a classification of maximal linear spaces of multipliers can be easily derived.)
Note that if the transformation $u\to \til u=\int\exp(A(u))du$ is inverted,
then $Q$ transforms to $\til Q= \exp(A(u))Q$.
(Also note that, under the restriction $Q=\phi(u)$ considered in \Ref{IbrTorTra10},
the classification reduces to just the first case with $m=\const$ and $Q=1$.)

\begin{table}[htb]
\centering
\caption{Adjoint-symmetries (``nonlinear self-adjointness'')}
\label{ex-table:adjsymm}
\begin{tabular}{c|ccc|l}
\hline
$Q(t,x,u)$ & $b(u)$ & $c(u)$ & $m(u)$ & conditions
\\
\hline
\hline
$e^{m_2t+m_3x}$
&
\arb
&
\arb
&
$m_1u + \int( m_2b + m_3c )du$
&
$m_1=m_3^2-m_2^2$
\\
\hline
$e^{\alpha x + \beta t}$
&
$b_0+b_1m'$
&
$c_0+c_1m'$
&
\arb
&
$\begin{aligned}
& b_1\beta +c_1\alpha =1
\\
& \beta(\beta -b_0) = \alpha(\alpha+c_0)
\end{aligned}$
\\
\hline
$e^{\gamma x} q(x\mp t)$
&
$b_0+b_1m'$
&
$c_0+c_1m'$
&
\arb
&
$\begin{aligned}
& \gamma=\pm b_0=-c_0,
\\
& b_1=1/b_0, c_1=-1/c_0,
\\
& q(\xi)=\text{ \arb}
\end{aligned}$
\\
\hline
\end{tabular}
\end{table}

The Fr\'echet  derivative operator
in the symmetry determining equation \eqref{symmdeteq}
and the adjoint of this operator
in the adjoint-symmetry determining equation \eqref{adjsymmdeteq}
are related by the integration-by-parts formula \eqref{adjoint}.
For a general DE system \eqref{sys},
this formula is given by
\begin{equation}\label{symm-adjsymm-ident}
Q^a(\delta_P F)_a - P^\alpha(\delta^*_Q F)_\alpha = D_i \Psi^i(P,Q;F)
\end{equation}
where the vector $\Psi^i(P,Q;F)$ is given by the explicit expression \eqref{frechetcurrent}
with $v=Q$, $w=P$, and $f=F$.
As shown in \Ref{Cav,Lun,AncBlu97},
this vector $\Psi^i(P,Q;F)$ will be a conserved current
\begin{equation}\label{symm-adjsymm-conslaw}
D_i \Psi^i(P,Q;F)|_\Esp = 0
\end{equation}
whenever the differential functions $P^\alpha$ and $Q^a$ respectively satisfy
the symmetry and adjoint-symmetry determining equations.
Moreover, it is straightforward to see
\begin{equation}\label{main_rel}
\Psi^i(P,Q;F) = \Phi^i(P;L)|_{v=Q}, 
\end{equation}
which follows from relation \eqref{current-rel},
where $\Phi^i(P;L)$ is the Noether conserved current \eqref{L-current}
and $L$ is the Lagrangian \eqref{L}.
Alternatively,
the equality \eqref{main_rel} can be derived indirectly
by applying formula \eqref{symm-adjsymm-ident}
to the variational identity \eqref{var_identity} with $v=Q$,
giving
\begin{equation}
\begin{aligned}
\pr\hat\X(L)  & = \phi^aF_a + v^a(\delta_P F)_a
= \phi^aF_a + P^\alpha(\delta_v^* F)_\alpha + D_i\Psi^i(P,v;F)
\\&
= \hat\X(v^a) E_{v^a}(L) + \hat\X(u^\alpha)E_{u^\alpha}(L) + D_i\Psi^i(P,v;F)
\end{aligned}
\end{equation}
which implies $\Psi^i(P,v;F)= \Phi^i(P;L)$ holds
(up to the possible addition of a total curl).

When the relation \eqref{main_rel} is combined with
Propositions~\ref{noetherformula} and~\ref{equiv-nonlin-self-adj},
the following main result is obtained.

\begin{thm}\label{equiv-thm}
For any DE system \eqref{sys} admitting an adjoint-symmetry \eqref{adjsymmdeteq}
(namely, a ``nonlinearly self-adjoint system'' in the general sense),
the conserved current \eqref{L-current} derived from
applying Noether's theorem to the extended Euler-Lagrange system \eqref{EL_sys}
using any given symmetry \eqref{hatXext}
is equivalent to the conserved current obtained using
the adjoint-symmetry/symmetry formula \eqref{symm-adjsymm-ident}.
\end{thm}

This theorem shows that the ``nonlinear self-adjointness'' method
based on Ibragimov's theorem
as developed in papers \cite{Ibr07,Ibr11,Gan11,Gan14,ZhaXie}
for DE systems with $M=m$
is just a special case of the adjoint-symmetry/symmetry formula \eqref{symm-adjsymm-ident}
introduced for general DE systems in prior papers \Ref{Cav,Lun,AncBlu97}
which were never cited.
Moreover,
the adjoint-symmetry/symmetry formula \eqref{symm-adjsymm-ident}
has the advantage that there is no need to extend the given DE system
by artificially adjoining variables to get an Euler-Lagrange system.

Another major advantage of the adjoint-symmetry/symmetry formula is that
it can be used to show how the resulting local conservation laws are, in general,
not necessarily non-trivial
and comprise only a subset of all of the non-trivial local conservation laws
admitted by a given DE system.
In particular,
in many applications of Theorem~\ref{Ibr-thm},
it is found that some non-trivial symmetries, particularly translation symmetries,
only yield trivial conservation laws
\cite{Fre13,Gan15,ZhaXie},
and that some local conservation laws are not produced even when all admitted symmetries are used.
These observations turn out to have a simple explanation
through the equivalence of Theorem~\ref{Ibr-thm}
and the adjoint-symmetry/symmetry formula \eqref{symm-adjsymm-ident},
as explained in the next section.

{\bf Example}:
For the semilinear wave equation \eqref{ex-eqn},
the extended Euler-Lagrange system in Ibragimov's notation
consists of
\begin{subequations}\label{ex-F-ext}
\begin{align}
& F=u_{tt} -u_{xx}+ b(u)u_t + c(u)u_x +m(u)
=\frac{\delta L}{\delta v}= 0 ,
\label{ex-F-ext-u-eqn}
\\
& F^*= v_{tt}  -v_{xx} -b v_t -c v_x +(b' u_t+c' u_x+m')v
=\frac{\delta L}{\delta u}= 0 , 
\label{ex-F-ext-v-eqn}
\end{align}
\end{subequations}
where $F^*$ is defined by the adjoint-symmetry equation \eqref{ex-adjsymmeq}
with $Q=v$,
and where the Lagrangian \eqref{L} is simply
$L=  vF= v(u_{tt} -u_{xx}+ b(u)u_t + c(u)u_x +m(u))$
in terms of the variables $u$ and $v$.
Consider any point symmetry of the wave equation \eqref{ex-F-ext-u-eqn} for $u$, 
given by a generator
\begin{equation}\label{ex-X}
\X = \tau(t,x,u)\parderop{t} + \xi(t,x,u)\parderop{x} + \eta(t,x,u)\parderop{u} .
\end{equation}
Its equivalent characteristic form is $\hat\X = P\parderop{u}$,
with $P=\eta-\tau u_t - \xi u_x$
satisfying the symmetry determining equation \eqref{ex-symmeq}
on the space of solutions $u(x)$ of the wave equation \eqref{ex-F-ext-u-eqn}.
Every point symmetry can be extended to a variational symmetry \eqref{Xext}
admitted by the Euler-Lagrange system,
which is given by the generator
$\X^\ext=\X + (\tau v_t+ \xi v_x - R_P^*(v))\parderop{v}$
where $R_P^*$ is the adjoint of the operator $R_P$
defined by relation \eqref{symmdeteq_offE} for the point symmetry
holding off of the solution space of the wave equation \eqref{ex-F-ext-u-eqn}.
In particular,
$R_P$ can be obtained by a straightforward computation of $\delta_P F = R_P(F)$,
where the terms in $\delta_P F$ are simplified
by using the equations $\tau_u=\xi_u=0$, $\tau_t=\xi_x$, and $\tau_x=\xi_t$
that arise from splitting the determining equation \eqref{ex-symmeq}.
This yields
\begin{equation}\label{ex-R_P}
R_P= -\tau D_t - \xi D_x +\eta_u -(\tau_t+\xi_x), 
\end{equation}
and thus
\begin{equation}\label{ex-adjR_P}
R_P^*= \tau D_t + \xi D_x +\eta_u .
\end{equation}
Hence, the variational symmetry is simply
\begin{equation}\label{ex-Xext}
\X^\ext=\tau\parderop{t} + \xi\parderop{x} + \eta\parderop{u} - \eta_u v\parderop{v}
\end{equation}
which is a point symmetry.

The action of this variational symmetry on the Lagrangian $L=vF$ is given by
\begin{equation}
\pr\X^\ext(L) = -\eta_u v F + v\pr\X(F) = -(\tau_t+\xi_x)vF
= -(D_t\tau +D_x\xi)L
\end{equation}
since $\pr\X(F) = \tau D_t F + \xi D_x F + R_P(F) = (\eta_u -(\tau_t+\xi_x))F$.
This symmetry action then can be combined with the variational identity \eqref{Xext_var_id}
to get the Noether relation
\begin{equation}\label{ex-noether_rel}
D_t(\tau L+\Phi^t(P;L)) + D_x(\tau L+\Phi^x(P;L))
= -\hat\X^\ext(v) F - \hat\X^\ext(u) F^*
\end{equation}
using $F=E_{v}(L)$ and $F^*=E_{u}(L)$,
where
\begin{equation}
\Phi^t(P;L) = vD_t P (b(u) v - v_t)P,
\quad
\Phi^x(P;L) = -vD_x P + (c(u) v +v_x)P
\end{equation}
are obtained from formula \eqref{L-current}.
This yields a conservation law
\begin{equation}\label{ex-L-conslaw}
(D_t C^t + D_x C^x)|_{\Esp(u,v)} =0,
\quad
C^t= \Phi^t(P;L) - \tau L,
\quad
C^x= \Phi^x(P;L)  - \xi L
\end{equation}
on the solution space $\Esp(u,v)$ of the Euler-Lagrange system $F=0$, $F^*=0$.
Since $L|_{\Esp(u,v)} =0$,
this conservation law is locally equivalent to the conservation law \eqref{uv_conslaw}
which is given by
\begin{equation}\label{ex-conslaw}
(D_t \hat C^t + D_x \hat C^x)|_{\Esp(u,v)} =0,
\quad
\hat C^t= \Phi^t(P;L),
\quad
\hat C^x= \Phi^x(P;L) .
\end{equation}
Moreover, from the identity \eqref{symm-adjsymm-ident}
relating the symmetry equation \eqref{ex-symmeq}
and the adjoint-symmetry equation \eqref{ex-adjsymmeq},
the conserved current $(\hat C^t,\hat C^x)$ in the conservation law \eqref{ex-conslaw}
is the same as the conserved current $(\Psi^t,\Psi^x)$ in the adjoint-symmetry/symmetry formula
\begin{equation}
\Psi^t(P,Q;F)|_{Q=v} = \Phi^t(P;L),
\quad
\Psi^x(P,Q;F)|_{Q=v}  = \Phi^x(P;L)
\end{equation}
where
\begin{equation}\label{ex-Psi}
\begin{aligned}
& \Psi^t(P,Q;F) = QD_t P + (b(u) Q - D_tQ)P ,
\\
& \Psi^x(P,Q;F) = -QD_x P + (c(u) Q+D_xQ)P .
\end{aligned}
\end{equation}

In \Ref{IbrTorTra10},
the conservation law formula \eqref{ex-L-conslaw} is used to obtain
a single local conservation law for a special case of the wave equation \eqref{ex-eqn}
given by $b=-c=-\ln(u)$ and $d=0$,
corresponding to $\til u_{tt}-\til u_{xx}-(\til u_t^2-\til u_x^2) +\til u(\til u_t-\til u_x)=0$
after an equivalence transformation $u\to \til u = e^{-u}$ is made.
The formula is applied to the adjoint-symmetry $\til Q=e^{-\til u}$
and the point symmetry $\widetilde\X=e^{(t+x)/2}\parderop{\til u}$
with characteristic $\til P=e^{(t+x)/2}$,
which respectively correspond to $Q=1$
and $\X= e^{(t+x)/2}u\parderop{u}$ with $P=e^{(t+x)/2}u$.
The likely reason why the obvious translation symmetries
$\widetilde\X=\parderop{t}$ and $\widetilde\X=\parderop{x}$
were not considered in \Ref{IbrTorTra10} 
is that these symmetries lead to locally trivial conservation laws
when $\til Q=e^{-\til u}$ is used.

To illustrate the situation,
consider the translation symmetries
\begin{equation}\label{ex-trans-symms}
\X_1 = \parderop{t},
\quad
\X_2 = \parderop{x}
\end{equation}
admitted by the wave equation \eqref{ex-eqn}
for arbitrary $b(u)$, $c(u)$, $m(u)$.
The characteristic functions of these two symmetries are, respectively, 
$P=-u_t$ and $P=-u_x$.
Local conservation laws can be obtained by applying the formula \eqref{ex-L-conslaw},
or its simpler equivalent version \eqref{ex-conslaw},
with $v=Q(t,x,u)$ being the adjoint-symmetries classified in table~\ref{ex-table:adjsymm}.
The resulting conserved currents $(\Psi^t,\Psi^x)$, modulo locally trivial currents,
are shown in table~\ref{ex-table:symm_adjsymm_conslaw}.

\begin{table}[htb]
\centering
\caption{Conserved currents from the adjoint-symmetry/symmetry formula}
\label{ex-table:symm_adjsymm_conslaw}
\begin{tabular}{c|c||c|c}
\hline
& & $\X=\parderop{t}$ & $\X=\parderop{x}$ \\\hline
conditions & $Q$ & $\Psi^t,\Psi^x$ & $\Psi^t,\Psi^x$
\\
\hline
\hline
$\begin{aligned}
& m=m_1u + m_2 \tint b\,du \\&\qquad
+ m_3\tint c\,du
\\
& m_1=m_3^2-m_2^2
\end{aligned}$
&
$e^{m_3x+m_2t}$
&
$\begin{aligned}
& m_2Q(u_t-m_2u+\tint b\,du),
\\
& m_2Q(m_3u-u_x+\tint c\,du)
\end{aligned}$
&
$\begin{aligned}
& m_3Q(u_t-m_2u+\tint b\,du),
\\
& m_3Q(m_3u-u_x+\tint c\,du)
\end{aligned}$
\\
\hline
$\begin{aligned}
& b=b_0+b_1m'
\\
& c=c_0+c_1m'
\\
& b_1\beta +c_1\alpha =1
\\
& \beta(\beta -b_0) = \alpha(\alpha+c_0)
\end{aligned}$
&
$e^{\alpha x + \beta t}$
&
$\begin{aligned}
& \beta Q(u_t-\beta u +\tint b\,du),
\\
& \beta Q(\alpha u-u_x+\tint c\,du)
\end{aligned}$
&
$\begin{aligned}
& \alpha Q(u_t-\beta u +\tint b\,du),
\\
& \alpha Q( \alpha u -u_x +\tint c\,du)
\end{aligned}$
\\
\hline
$\begin{aligned}
& b=\pm(\gamma +\tfrac{1}{\gamma} m')
\\
& c=-\gamma +\tfrac{1}{\gamma} m'
\end{aligned}$
&
$e^{\gamma x} q(x\mp t)$
&
$\begin{aligned}
& -e^{\gamma x}( q'' u 
\pm q'(u_t +\tint b\,du) ),
\\
& \pm e^{\gamma x}( (\gamma q'-q'')u  \\&
+ q(u_x \mp\tint b\,du) )
\end{aligned}$
&
$\begin{aligned}
& e^{\gamma x}( \pm (q'' +\gamma q')u \\&
+(q'+\gamma q)(u_t +\tint b\,du) ),
\\
& e^{\gamma x}( (q'' -\gamma^2 q)u \\&
{-}(q'+\gamma q)(u_x\mp\tint b\,du) )
\end{aligned}$
\\
\hline
\end{tabular}
\end{table}

Notice that for $Q=\const$ the conserved currents $(\Psi^t,\Psi^x)$
obtained from the two translation symmetries vanish.
This implies that Ibragimov's theorem \eqref{ex-L-conslaw}
yields just trivial conserved currents $(\Psi^t,\Psi^x)$
for some cases of the wave equation \eqref{ex-eqn}
when a non-trivial conserved current exists. 
A full explanation of why this occurs will be given in the next section.

\section{Properties of conservation laws generated by the adjoint-symmetry/symmetry formula and Ibragimov's theorem}
\label{sec:symmaction}

To determine when a conserved current is locally trivial,
or when two conserved currents are locally equivalent,
it is useful to have a characteristic (canonical) form for local conservation laws,
in analogy to the characteristic form for local symmetries.

Any local conservation law \eqref{conslaw}
can be expressed as a divergence identity \cite{Olv}
\begin{equation}\label{conslaw_offE}
D_i C^i = R_C{}^{a} F_a + R_C{}^{a i} D_{i}F_a + \cdots + R_C{}^{a i_1\cdots i_r} D_{i_1}\cdots D_{i_r} F_a
\end{equation}
by moving off of the solution space $\Esp$ of the system,
where $R_C{}^{a},R_C{}^{a i},\ldots,R_C{}^{a i_1\cdots i_r}_{a}$
are some differential functions that are non-singular on $\Esp$,
under some mild regularity conditions \cite{Anc-review} on the form of the DEs \eqref{sys}.
Integration by parts on the terms on the right-hand side in this identity \eqref{conslaw_offE}
then yields
\begin{equation}\label{chareqn}
D_i \tilde C^i= Q_C^a F_a
\end{equation}
with
\begin{equation}\label{multr}
Q_C^a = R_C{}^{a} - D_i R_C{}^{a i} + \cdots + (-1)^r D_{i_1}\cdots D_{i_r} R_C{}^{a i_1\cdots i_r}, 
\end{equation}
where
\begin{equation}
\tilde C^i|_\Esp = C^i|_\Esp
\end{equation}
reduces to the conserved vector in the given conservation law \eqref{conslaw}.
Hence,
\begin{equation}
(D_i \tilde C^i)|_\Esp= 0
\end{equation}
is a locally equivalent conservation law.
The identity \eqref{chareqn} is called the \emph{characteristic equation} \cite{Olv}
for the conservation law \eqref{conslaw},
and the set of differential functions \eqref{multr}
is called the {\em conservation law multiplier} \cite{Olv}.
In general
a set of functions $f^a(t,x,u,\p u,\p^2 u,\ldots\p^s u)$ will be a multiplier
iff it is non-singular on $\Esp$ and its summed product with the DEs $F_a$ in the system
has the form of a total divergence.

For a given local conservation law,
the multiplier arising from the integration by parts formula \eqref{multr}
will be unique iff the coefficient functions in the characteristic equation \eqref{conslaw_offE}
are uniquely determined by the conserved vector $C^i$.
This uniqueness holds straightforwardly for any DE system consisting of
a single equation that can be expressed in a solved form for a leading derivative
\cite{Anc-review}. 
For DE systems containing more than one equation,
some additional technical requirements are necessary \cite{Olv}.
In particular,
it is necessary that a DE system have no differential identities \cite{Olv},
and it is sufficient that a DE system have a generalized Cauchy-Kovalevskaya form
\cite{Olv,Mar-Alo,AncBlu02a,AncBlu02b}.
A concrete necessary and sufficient condition,
which leads to the following uniqueness result,
is stated in \Ref{Anc-review}.

\begin{prop}\label{correspondence-no-diffids}
For any closed DE system \eqref{sys}
having a solved form in terms of leading derivatives and having no differential identities,
a conserved current is locally trivial \eqref{trivconslaw}
iff its corresponding multiplier \eqref{multr} vanishes
when evaluated on the solution space of the system.
\end{prop}

This class of DE systems includes nearly all systems of physical interest,
apart from systems such as the Maxwell equations and the incompressible fluid equations,
which possess differential identities.
Often the distinction between systems with and without differential identities is overlooked
in the literature on conservation law multipliers.

The importance of Proposition~\ref{correspondence-no-diffids} is that
in a wide class of DE systems it establishes that
a unique characteristic form for locally equivalent conservation laws
is provided by multipliers.
From this result,
it is now straightforward to derive a simple condition to detect
when a local conservation law given by the adjoint-symmetry/symmetry formula \eqref{symm-adjsymm-ident}
is locally trivial \eqref{trivconslaw}.

Let $P^\alpha(x,u,\p u,\ldots,\p^r u)$ be the characteristic functions
defining a symmetry \eqref{symmdeteq},
and let $Q^a(x,u,\p u,\ldots,\p^s u)$ be a set of differential functions
defining an adjoint-symmetry \eqref{adjsymmdeteq}.
Then the adjoint-symmetry/symmetry formula \eqref{symm-adjsymm-ident}
yields a local conservation law \eqref{symm-adjsymm-conslaw}.
The characteristic equation of this conservation law is given by
substituting the symmetry identity \eqref{symmdeteq_offE}
and the adjoint-symmetry identity \eqref{adjsymmdeteq_offE}
into the formula \eqref{symm-adjsymm-ident}
to get
\begin{equation}
D_i \Psi^i(P,Q;F) = Q^a R_P(F)_a - P^\alpha R_Q(F)_\alpha  .
\end{equation}
Integration by parts gives
\begin{equation}\label{chareqn-symm-adjsymm}
D_i \tilde\Psi^i(P,Q;F) = (R_P^*(Q)^a  - R_Q^*(P)^a)F_a
\end{equation}
where
\begin{equation}
\tilde \Psi^i(P,Q;F)|_\Esp = \Psi^i(P,Q;F)|_\Esp .
\end{equation}
Hence, the conservation law multiplier is given by \cite{AncKar}
\begin{equation}\label{multr-symm-adjsymm}
Q_\Psi^a = R_P^*(Q)^a  - R_Q^*(P)^a .
\end{equation}
This yields the following result.

\begin{prop}\label{adjsymm-symm-trivial}
The adjoint-symmetry/symmetry formula \eqref{symm-adjsymm-ident}
for a given DE system \eqref{sys}
produces a locally trivial conservation law if the condition
\begin{equation}\label{triv-multr-adjsymm-symm}
(R_P^*(Q)^a  - R_Q^*(P)^a)|_\Esp =0
\end{equation}
holds for the given symmetry and adjoint-symmetry pair,
where
$P^\alpha(x,u,\p u,\ldots,\p^r u)$ is the set of characteristic functions of the symmetry \eqref{symmdeteq}
and $Q^a(x,u,\p u,\ldots,\p^s u)$ is the set of functions defining the adjoint-symmetry \eqref{adjsymmdeteq}.
This condition \eqref{triv-multr-adjsymm-symm} is also sufficient
whenever the DE system \eqref{sys} belongs to the class
stated in Proposition~\ref{correspondence-no-diffids}.
\end{prop}

Through the equivalence stated in Theorem~\ref{equiv-thm},
which relates the adjoint-symmetry/symmetry formula \eqref{symm-adjsymm-ident}
and the generalized version of Ibragimov's conservation law formula in Theorem~\ref{Ibr-thm},
it follows that ``nonlinear self-adjointness'' through a differential substitution with $v=Q$
produces a conservation law \eqref{L-current} that is locally trivial
when $Q^a(x,u,\p u,\ldots,\p^s u)$ and $P^\alpha(x,u,\p u,\ldots,\p^r u)$
satisfy condition \eqref{triv-multr-adjsymm-symm}.

A useful remark is that
the triviality condition \eqref{triv-multr-adjsymm-symm} can be checked directly,
without the need to derive the local conservation law itself.

{\bf Example}:
For the semilinear wave equation \eqref{ex-eqn},
consider the conserved currents obtained in table~\ref{ex-table:symm_adjsymm_conslaw},
which are generated from the three adjoint-symmetries
\begin{equation}\label{ex-adjsymms}
Q_1=e^{m_2t+m_3x},
\quad
Q_2=e^{\alpha x + \beta t},
\quad
Q_3=e^{\gamma x} q(x\mp t), 
\end{equation}
and the two translation symmetries \eqref{ex-trans-symms}.
The operators $\delta_P F = R_P(F)$ associated to the characteristics
\begin{equation}\label{ex-symmchars}
P_1 = -u_t,
\quad
P_2 = -u_x
\end{equation}
of these two symmetries \eqref{ex-trans-symms}
are given by the formula \eqref{ex-R_P},
which yields
\begin{equation}\label{ex-symms-R}
R_{P_1}=-D_t,
\quad
R_{P_2}=-D_x .
\end{equation}
For adjoint-symmetries of the form $Q(t,x,u)$,
the operator $\delta_Q^* F = R_Q(F)$ is easily found to be
\begin{equation}\label{ex-R_Q}
R_Q= Q_u .
\end{equation}
Hence,
the operators associated to the three adjoint-symmetries \eqref{ex-adjsymms}
are simply
\begin{equation}\label{ex-adjsymms-R}
R_{Q_1}=R_{Q_2}=R_{Q_3}=0 .
\end{equation}
The triviality condition \eqref{triv-multr-adjsymm-symm} is then given by
\begin{equation}\label{ex-triv-multr-adjsymm-symm}
R_{P_1}^*(Q_{l})  - R_{Q_l}^*(P_1) = D_t Q_{l} =0,
\quad
R_{P_2}^*(Q_{l})  - R_{Q_l}^*(P_2) = D_x Q_{l} =0,
\quad
l=1,2,3 .
\end{equation}
This shows that the two conserved currents obtained from $Q_1$
will be trivial when $m_2=0$ and $m_3=0$ hold, respectively,
and that likewise the two conserved currents obtained from $Q_2$
will be trivial when $\beta=0$ and $\alpha=0$ hold, respectively.
Similarly, for $Q_3$,
the first conserved current will be trivial when $q'=0$ holds,
while the second conserved current will be trivial when $q'+\gamma q=0$ holds,
corresponding to $q= e^{-\gamma(x\pm t)}$.
These trivial cases can be seen to occur directly from the explicit expressions
for the conserved currents  $(\Psi^t,\Psi^x)$ in table~\ref{ex-table:symm_adjsymm_conslaw}.

In general,
while the adjoint-symmetry/symmetry formula \eqref{symm-adjsymm-ident}
(and hence Ibragimov's theorem) looks very appealing,
it has major drawbacks that in many examples
\cite{IbrTorTra10,IbrTorTra11,Gan11,FreSam12,Gan14,Zha,ZhaXie}
the selection of a symmetry must be fitted to the form of the adjoint-symmetry
to produce a non-trivial conservation law,
and that in other examples \cite{Fre13,Gan15,ZhaXie}
no non-trivial conservation laws are produced
when only translation symmetries are available.
More importantly,
it is \emph{not} (as is sometimes claimed)
a generalization of Noether's theorem to non-variational DE systems.

As a reinforcement of these statements,
consider the situation of variational DE systems,
where adjoint-symmetries coincide with symmetries.
Then the adjoint-symmetry/symmetry formula \eqref{symm-adjsymm-ident}
produces a conserved current directly from any pair of symmetries
admitted by a given variational DE system.
But, from Noether's theorem,
this conserved current must also arise directly
from some variational symmetry of the system.
Moreover, if the pair of symmetries being used are variational symmetries that happen to commute with each other,
then the resulting conserved current turns out to be trivial
as shown in \Ref{AncBlu96}.

To understand these aspects and other properties of the formula,
the determining equations for multipliers are needed.

\subsection{Multiplier determining equations}

All conservation law multipliers for any given DE system can be determined from
the property \eqref{eulerop-kernel}--\eqref{eulerop-div} that
a differential function is a total divergence iff it is annihilated by the Euler operator \eqref{eulerop}.
Specifically,
when this property is applied directly to the characteristic equation \eqref{chareqn}
for local conservation laws,
it yields the determining equations
\begin{equation}\label{eulerop-multr-eqn}
E_{u^\alpha}(Q_C^aF_a) =0,
\quad
\alpha = 1,\ldots,m
\end{equation}
which are necessary and sufficient \cite{Olv}
for a set of differential functions $Q_C^a(x,u,\p u,\ldots,\p^s u)$ to be a multiplier
for a local conservation law \eqref{conslaw}.
Note these equations \eqref{eulerop-multr-eqn} must hold identically in jet space
(and not just on the solution space $\Esp$ of the DE system).

The multiplier determining equations \eqref{eulerop-multr-eqn} have a close connection
to the determining equation \eqref{adjsymmdeteq} for adjoint-symmetries.
This can be immediately seen from the product rule \eqref{eulerop-productrule}
obeyed by the Euler operator,
which gives
\begin{equation}\label{multrdeteq}
0=E_{u^\alpha}(Q_C^aF_a) = (\delta^*_{Q_C} F)_\alpha + (\delta^*_F Q_C)_\alpha
\end{equation}
holding identically in jet space $J(x,u,\p u,\p^2 u,\ldots)$.
Notice if this equation \eqref{multrdeteq} is restricted to the solution space
$\Esp\subset J$ of the given DE system \eqref{sys},
then it coincides with the adjoint-symmetry determining equation \eqref{adjsymmdeteq}.
Hence, every conservation law multiplier is an adjoint-symmetry.
This is a well-known result
\cite{Olv,AncBlu97,AncBlu02a,AncBlu02b,2ndbook,Anc-review}.
What is not so well-known are the other conditions \cite{AncBlu97,Anc-review}
that an adjoint-symmetry must satisfy to be a conservation law multiplier.
These conditions arise from splitting the determining equation \eqref{multrdeteq}
with respect to $F_a$ and its total derivatives.
As shown in \Ref{Anc-review},
the splitting can be derived by using the adjoint-symmetry identity \eqref{adjsymmdeteq_offE}
combined with the expression
\begin{equation}
(\delta^*_F Q)_\alpha =
F_a\parder{Q^a}{u^\alpha} -D_i\Big(F_a\parder{Q^a}{u^\alpha_i}\Big) +\cdots +(-1)^s D_{i_1}\cdots D_{i_s}\Big(F_a\parder{Q^a}{u^\alpha_{i_1\cdots i_s}}\Big) .
\end{equation}
Then, in the determining equation \eqref{multrdeteq},
the coefficients of $F_a$, $D_i F_a$, and so on
yield the system of equations \cite{Anc-review}
\begin{equation}\label{multr-adjsymmeq}
(\delta^*_{Q_C} F)_\alpha|_\Esp =0
\end{equation}
and
\begin{subequations}\label{multr-helmholtzeq}
\begin{gather}
R_Q{}^{a}_{\alpha} + E_{u^\alpha}(Q_C^a) =0
\label{multr-helmholtz-0th}
\\
R_Q{}^{a i_1\cdots i_q}_{\alpha} +(-1)^q E_{u^\alpha}^{(i_1\cdots i_q)}(Q_C^a) =0,
\quad
q=1,\ldots,s
\label{multr-helmholtz-higher}
\end{gather}
\end{subequations}
where
$R_Q{}^{a}_{\alpha}$ and $R_Q{}^{a i_1\cdots i_q}_{\alpha}$
are the coefficient functions of the linear differential operator \eqref{adjsymm_op}
determined by equation \eqref{multr-adjsymmeq},
and where
$E_{u^\alpha}$ is the Euler operator \eqref{eulerop}
and
$E_{u^\alpha}^{(i_1\cdots i_q)}$
is a higher-order Euler operator defined by \cite{Olv,Anc-review}
\begin{equation}\label{highereulerop}
\begin{aligned}
E_{u^\alpha}^{(i_1\cdots i_q)}(f) & =
\parder{f}{u^\alpha_{i_1\cdots i_q}} 
-\tbinom{q+1}{1}D_i\Big(\parder{f}{u^\alpha_{i_1\cdots i_qi} }\Big) + \cdots
\\&\qquad
+ (-1)^r \tbinom{q+r}{r}D_{i_1}\cdots D_{i_r}\Big(\parder{f}{u^\alpha_{i_1\cdots i_qi_{q+1}\cdots i_{q+r}}}\Big),
\quad
q=1,2,\ldots
\end{aligned}
\end{equation}
for an arbitrary differential function $f(x,u,\p u,\ldots,\p^s u)$.
This system \eqref{multr-adjsymmeq}--\eqref{multr-helmholtzeq}
constitutes a determining system for conservation law multipliers.
Its derivation requires the same technical conditions on the form of the DE system \eqref{sys}
as stated in Proposition~\ref{correspondence-no-diffids}.

\begin{thm}\label{multrdetsys}
The determining equation \eqref{multrdeteq} for conservation law multipliers
of a general DE system \eqref{sys}
is equivalent to the linear system of equations \eqref{multr-adjsymmeq}--\eqref{multr-helmholtzeq}.
In particular,
multipliers are adjoint-symmetries \eqref{multr-adjsymmeq}
satisfying Helmholtz-type conditions \eqref{multr-helmholtzeq}
which are necessary and sufficient for an adjoint-symmetry
to have the variational form \eqref{multr} derived from a conserved current.
\end{thm}

This well-know result \cite{AncBlu97,AncBlu02a,AncBlu02b,Anc-review}
gives a precise relationship between adjoint-symmetries and multipliers,
or equivalently between ``nonlinear self-adjointness'' and multipliers.
In particular,
it provides necessary and sufficient conditions for an adjoint-symmetry to be a multiplier.
The simplest situation is when adjoint-symmetries of the lowest-order form $Q^\alpha(x,u)$ are considered,
which corresponds to ``nonlinear self-adjointness'' without differential substitutions.
In this case, the only condition is equation \eqref{multr-helmholtz-0th},
which reduces to
$R_Q{}^{a}_{\alpha} + \dfrac{\p Q^a}{\p u^\alpha}=0$.
This condition is, in general, non-trivial.
(Unfortunately, some recent work \cite{Zha} incorrectly asserts that, for any DE system,
every adjoint-symmetry of the form $Q^\alpha(x,u)$ is a multiplier.)

When Theorem~\ref{multrdetsys} is applied to variational DE systems,
it yields the following well-known connection \cite{AncBlu97,AncBlu02a,AncBlu02b,Anc-review}
with Noether's theorem.

\begin{cor}
For a variational DE system \eqref{variational_sys},
the multiplier determining system \eqref{multr-adjsymmeq}--\eqref{multr-helmholtzeq}
reduces to a determining system for variational symmetries.
In particular,
the determining equation for adjoint-symmetries \eqref{multr-adjsymmeq}
coincides with the determining equation for symmetries \eqref{symmdeteq},
and the Helmholtz-type conditions \eqref{multr-helmholtzeq}
coincide with the necessary and sufficient conditions for a symmetry to be variational
(namely, that $\pr\hat\X(L) = D_i\Gamma^i$ holds
for some differential vector function $\Gamma^i$,
where $L$ is the Lagrangian \eqref{variational_sys_L}).
\end{cor}

Note that, in this modern formulation of Noether's theorem,
the use of a Lagrangian is completely by-passed through
the Helmholtz-type conditions \eqref{multr-helmholtzeq}.

{\bf Example}:
For the semilinear wave equation \eqref{ex-eqn},
the determining equation for multipliers of lowest-order form $Q_C(t,x,u)$ is given by
\begin{equation}\label{ex-multreq}
0=E_{u}(Q_C F) = \delta^*_{Q_C} F + \delta^*_F Q_C .
\end{equation}
Since $Q_C(t,x,u)$ does not depend on derivatives of $u$,
this determining equation splits with respect to the variables $u_t,u_x,u_{tt},u_{xx}$,
giving an overdetermined linear system
which can be derived and solved directly by Maple.
This provides the simplest computational route to finding all multipliers of lowest-order form.
The connection between multipliers and adjoint-symmetries arises
when the determining equation \eqref{ex-multreq} is instead split into the two terms
$\delta^*_{Q_C} F$ and $\delta^*_F Q_C$,
which are given by
\begin{align}
& \delta^*_{Q_C} F = D_t^2Q_C  -D_x^2Q_C -b D_tQ_C -c D_x Q_C +(b' u_t+c' u_x+m')Q_C
= R_{Q_C}(F)
\label{ex-split1}
\\
& \delta^*_F Q_C = \parder{Q_C}{u}F = E_u(Q_C)F
\label{ex-split2}
\end{align}
where the operator $R_{Q_C}$ is obtained from expression \eqref{ex-R_Q}.
Hence,
on the solution space $\Esp$ of the wave equation \eqref{ex-eqn},
the multiplier determining equation reduces to the adjoint-symmetry equation \eqref{ex-adjsymmeq}.
Off of the solution space $\Esp$,
the multiplier determining equation then becomes
\begin{equation}
0= R_{Q_C}(F) +E_u(Q_C)F = 2\parder{Q_C}{u}F
\end{equation}
which splits with respect to $F$, yielding
\begin{equation}\label{ex-multr-helmholtzeq}
\parder{Q_C}{u} =0 .
\end{equation}
This Helmholtz-type equation \eqref{ex-multr-helmholtzeq}
together with the adjoint-symmetry equation \eqref{ex-adjsymmeq}
constitutes the determining system \eqref{multr-adjsymmeq}--\eqref{multr-helmholtzeq}
for finding all lowest-order multipliers $Q_C(t,x,u)$
admitted by the wave equation \eqref{ex-eqn}.

The Helmholtz-type equation \eqref{ex-multr-helmholtzeq} directly shows that
all adjoint-symmetries of the form $Q(t,x)$ are conservation laws multipliers $Q_C(t,x)$,
and so the three adjoint-symmetries \eqref{ex-adjsymms}
each determine a non-trivial conserved current through the characteristic equation
\begin{equation}\label{ex-chareq}
Q_C F = D_t \hat C^t + D_x \hat C^x .
\end{equation}
These conserved currents $(\hat C^t,\hat C^x)$ can be derived
in terms of the multipliers $Q_C(t,x,u)$ in several different ways.
One simple way is by applying integration by parts to the terms in $Q_CF$
to get a total time derivative $D_t \hat C^t$ plus a total space derivative $D_x \hat C^x$,
which yields $(\hat C^t,\hat C^x)$.
Another way is by taking $\hat C^t(t,x,u,u_t,u_x)$ and $\hat C^x(t,x,u,u_t,u_x)$
as unknowns and splitting the characteristic equation with respect to
$u_{tt}$, $u_{tx}$, $u_{xx}$ to get a linear system of determining equations
that can be integrated.
The resulting conserved currents are shown in table~\ref{ex-table:conslaw}.
The specific relationship between these conserved currents
and the conserved currents derived in table~\ref{ex-table:symm_adjsymm_conslaw}
will be explained in the next subsection.

\begin{table}[htb]
\centering
\caption{Conserved currents}
\label{ex-table:conslaw}
\begin{tabular}{c|c||c|c}
\hline
conditions & $Q_C$ & $\hat C^t$ & $\hat C^x$
\\
\hline
\hline
$\begin{aligned}
& m=m_1u + m_2 \tint b\,du \\&\qquad
+ m_3\tint c\,du
\\
& m_1=m_3^2-m_2^2
\end{aligned}$
&
$e^{m_3x+m_2t}$
&
$e^{m_3x+m_2t}(u_t-m_2u+\tint b\,du)$
&
$e^{m_3x+m_2t}(m_3u-u_x+\tint c\,du)$
\\
\hline
$\begin{aligned}
& b=b_0+b_1m'
\\
& c=c_0+c_1m'
\\
& b_1\beta +c_1\alpha =1
\\
& \beta(\beta -b_0) = \alpha(\alpha+c_0)
\end{aligned}$
&
$e^{\alpha x + \beta t}$
&
$e^{\alpha x + \beta t}(u_t-\beta u +\tint b\,du)$
&
$e^{\alpha x + \beta t}(\alpha u-u_x+\tint c\,du)$
\\
\hline
$\begin{aligned}
& b=\pm(\gamma +\tfrac{1}{\gamma} m')
\\
& c=-\gamma +\tfrac{1}{\gamma} m'
\end{aligned}$
&
$e^{\gamma x} q(x\mp t)$
&
$e^{\gamma x}( q(u_t +\tint b\,du)  \pm q'u )$
&
$e^{\gamma x}( (q' -\gamma q)u -(u_x \mp \tint b\,du) )$
\\
\hline
\end{tabular}
\end{table}

\subsection{Conservation laws produced by a multiplier/symmetry pair}

From Theorem~\ref{multrdetsys} and Proposition~\ref{correspondence-no-diffids},
every multiplier admitted by a given DE system determines,
up to local equivalence, a conserved current for the system.
Since multipliers are adjoint-symmetries,
the adjoint-symmetry/symmetry formula \eqref{symm-adjsymm-ident}
can be applied by using any multiplier \eqref{multr} together with any symmetry \eqref{symm}.
The resulting conserved current produced this way is given by
\begin{equation}\label{symm-multr-ident}
Q_C^a(\delta_P F)_a - P^\alpha(\delta^*_{Q_C} F)_\alpha  = D_i \Psi^i(P,Q_C;F)
\end{equation}
where $P^\alpha(x,u,\p u,\ldots,\p^r u)$ is a given symmetry characteristic
and $Q_C^a(x,u,\p u,\ldots,\p^s u)$ is a given multiplier.
The following result characterizing these conserved currents
will now be established for DE systems
in the class stated in Proposition~\ref{correspondence-no-diffids}.
The case of DE systems consisting of a single DE has appeared previously in \Ref{Anc16}.

\begin{thm}\label{symmaction}
Let $\Psi^i(P,Q_C;F)$ be the conserved current produced from
the adjoint-symmetry/symmetry formula \eqref{symm-adjsymm-ident}
by using any multiplier $Q_C^a(x,u,\p u,\ldots,\p^s u)$
together with any symmetry characteristic $P^\alpha(x,u,\p u,\ldots,\p^r u)$.
This conserved current $\Psi^i(P,Q_C;F)$ is locally equivalent to
a conserved current \eqref{symm_on_C} that is given by
the infinitesimal action of the symmetry $\hat\X_P= P^\alpha\parderop{u^\alpha}$
applied to the conserved current $C^i$ determined by the multiplier $Q_C^a$.
In particular, $\Psi^i(P,Q_C;F)$ and $C^i$ are related by
\begin{equation}
(\Psi^i(P,Q_C;F) - \pr\hat\X_P(C^i))|_\Esp = D_j\Theta^{ij}
\end{equation}
for some differential antisymmetric tensor function
$\Theta^{ij}(x,u,\p u,\ldots,\p^k u)$.
\end{thm}

The proof consists of showing that both conserved currents
$\Psi^i(P,Q_C;F)$ and $\pr\hat\X_P(C^i)$ have the same multiplier.
Consider the local conservation law determined by the multiplier $Q_C^a$.
The symmetry $\hat\X_P$ applied to the characteristic equation \eqref{chareqn} of this conservation law
yields
\begin{equation}\label{X_on_chareqn}
\pr\hat\X_P(D_i\tilde C^i) = \pr\hat\X(Q_C^aF_a)
= \delta_P(Q_C^aF_a) = (\delta_P Q_C)^a F_a + Q_C^a(\delta_P F_a) .
\end{equation}
The second term in this equation can be expressed as
\begin{equation}\label{X_on_chareqn_term2}
Q_C^a(\delta_P F_a) = Q_C^a R_P(F)_a = F_a R_P^*(Q_C)^a + D_i \Gamma^i(Q_C,F;P)
\end{equation}
using the symmetry identity \eqref{symmdeteq_offE}
combined with integration by parts,
where $\Gamma^i(Q_C,F;P)|_\Esp =0$.
Next,
the first term in equation \eqref{X_on_chareqn} can be expressed as
\begin{equation}
\begin{aligned}
(\delta_P Q_C)^a F_a  & = P^\alpha (\delta^*_F Q_C)_\alpha + D_i \Psi^i(P,Q_C;F)
= -P^\alpha (\delta^*_{Q_C}F)_\alpha + D_i \Psi^i(P,Q_C;F)
\\&
= -P^\alpha R_{Q_C}(F)_\alpha + D_i \Psi^i(P,Q_c;F)
\end{aligned}
\end{equation}
through the Fr\'echet  derivative identity \eqref{adjoint}
combined with the multiplier determining equation \eqref{multrdeteq}
and the adjoint-symmetry identity \eqref{adjsymmdeteq_offE}.
Integration by parts then yields
\begin{equation}\label{X_on_chareqn_term1}
(\delta_P Q_C)^a F_a  = -F_a R^*_{Q_C}(P)^a + D_i(\Psi^i -\Gamma^i(P,F;Q_C))
\end{equation}
where $\Gamma^i(P,F;Q_C)|_\Esp =0$.
Substitution of expressions \eqref{X_on_chareqn_term1} and \eqref{X_on_chareqn_term2}
into equation \eqref{X_on_chareqn} gives
\begin{equation}\label{X_on_chareqn_terms}
\pr\hat\X_P(D_i\tilde C^i) = ( R_P^*(Q_C)^a  - R^*_{Q_C}(P)^a ) F_a
+ D_i(\Psi^i(P,Q_c;F) +\Gamma^i(Q_C,F;P)-\Gamma^i(P,F;Q_x)) .
\end{equation}
Finally, since $\pr\hat\X_P$ commutes with total derivatives \cite{Olv,Anc-review},
this yields
\begin{equation}
D_i(\pr\hat\X_P(C^i) + \tilde\Gamma^i) = Q_\Psi^a F_a
\end{equation}
where $\tilde\Gamma^i|_\Esp=0$ is a locally trivial conserved current,
and where
\begin{equation}\label{multr-symm-multr}
Q_\Psi^a = R_P^*(Q_C)^a  - R_Q^*(P)^a
\end{equation}
is the multiplier \eqref{multr-symm-adjsymm} of the local conservation law \eqref{chareqn-symm-adjsymm}
from the adjoint-symmetry/symmetry formula \eqref{symm-adjsymm-ident}
with $Q^a= Q_C^a$.
This completes the proof.

Theorem~\ref{symmaction} is a generalization of a similar result \cite{Cav,Lun,AncBlu96}
for variational DE systems,
where the adjoint-symmetry/symmetry formula \eqref{symm-adjsymm-ident}
reduces to a formula using any pair of symmetries.

\begin{cor}\label{symmaction_variational}
For a variational DE system,
let $\Psi^i(P,Q_C;F)$ be the conserved current produced from
the adjoint-symmetry/symmetry formula \eqref{symm-adjsymm-ident}
by using any symmetry characteristic $P^\alpha(x,u,\p u,\ldots,\p^r u)$
together with any multiplier $Q_C^\alpha(x,u,\p u,\ldots,\p^s u)$
given by a variational symmetry characteristic.
The conserved current $\Psi^i(P,Q_C;F)$ is locally equivalent to
a conserved current \eqref{symm_on_C} that is given by
the infinitesimal action of the symmetry $\hat\X_P = P^\alpha\parderop{u^\alpha}$
applied to the conserved current $C^i$ determined by the multiplier $Q_C^\alpha$.
Moreover, through Noether's theorem,
the multiplier of this conserved current $\Psi^i(P,Q_C;F)$
is the characteristic of a variational symmetry given by the commutator
of the symmetries $\hat\X_P = P^\alpha\parderop{u^\alpha}$
and $\hat\X_{Q_C} = Q_C^\alpha\parderop{u^\alpha}$.
\end{cor}

Several basic properties of the adjoint-symmetry/symmetry formula \eqref{symm-adjsymm-ident}
can be deduced from Theorem~\ref{symmaction},
as first shown in \Ref{Anc16} for DE systems consisting of a single DE.

\begin{thm}\label{PQformula}
(i) For a given DE  system \eqref{sys},
let $Q_C^a$ be the multiplier for a local conservation law
in which the components of the conserved current $C^i$
have no explicit dependence on $x$.
Then, using any translation symmetry $\X=a^i\parderop{x^i}$,
with characteristic $P^\alpha=-a^i u^\alpha_i$ where $a^i$ is a constant vector,
the conserved current $\Psi^i(P,Q_C;F)$ is locally trivial.
(ii) For a given DE system \eqref{sys} that possesses a scaling symmetry
$\X=a_{(i)} x^i\parderop{x^i}+b_{(\alpha)} u^\alpha\parderop{u^\alpha}$, 
where  $a_{(i)},b_{(\alpha)}$ are constants,
let $Q_C^a$ be the multiplier for a local conservation law
in which the components of the conserved current $C^i$ are scaling homogeneous.
Then, using the characteristic $P^\alpha=b_{(\alpha)} u^\alpha -a_{(i)} x^i u^\alpha_i$ of the scaling symmetry,
the conserved current $\Psi^i(P,Q_C;F)$ is locally equivalent to
a multiple $w$ of the conserved current $C^i$ determined by $Q_C^a$.
This multiple, $w=\const$, is the scaling weight of the conserved integral given by
$\int_{\p\Omega} C^i dS_i$
where $\Omega$ is any closed domain in $\Rnum^n$
and $\p\Omega$ is its boundary surface.
\end{thm}

The proof is a straightforward extension of the proof in \Ref{Anc16} and will be omitted.

Part (i) of this theorem explains the observations made in many recent papers
in which Ibragimov's theorem gave only trivial local conservation laws.
This will happen whenever the only local symmetries admitted by a DE system
are translations and the only admitted adjoint-symmetries have no dependence on $x$.

Part (ii) of the theorem first appeared in \Ref{Anc03}.
It shows that the local conservation laws admitted by any DE system with a scaling symmetry
can be obtained from an algebraic formula using the conservation law multipliers.
This explains why in many recent papers the use of scaling symmetries
in Ibragimov's theorem has produced non-trivial local conservation laws.

A more important point comes from putting together Theorem~\ref{symmaction}
and Theorem~\ref{multrdetsys}.
Together, these two theorems show that the adjoint-symmetry/symmetry formula \eqref{symm-adjsymm-ident}
cannot produce any ``new'' local conservation laws,
since any local conservation law admitted by a given DE system
must already arise directly from a multiplier.
Moreover,
for this formula to generate all of the local conservation laws for a given DE system,
it seems plausible that the set of admitted symmetries needs to act transitively
on set of admitted local conservation laws,
so then every multiplier arises from some symmetry applied to some multiplier.
The need for a transitive action is especially clear from Corollary~\ref{symmaction_variational},
since if a pair of commuting variational symmetries is used in the formula,
then the resulting local conservation law will have a trivial multiplier
and hence will be a locally trivial conservation law.

These significant deficiencies should discourage
the unnecessary use of the adjoint-symmetry/symmetry formula \eqref{symm-adjsymm-ident}
--- and consequently the unnecessary use of Ibragimov's theorem ---
when local conservation laws are being sought for a given DE system.
It is much simpler and more direct to find all multipliers
and then derive the conserved currents determined by these multipliers,
as will be explained further in the next section.

{\bf Example}:
For the semilinear wave equation \eqref{ex-eqn},
table~\ref{ex-table:symm_adjsymm_conslaw} shows the conserved currents
obtained from the adjoint-symmetry/symmetry formula \eqref{symm-adjsymm-ident}.
Each of these conserved currents can be checked
to satisfy the characteristic equation \eqref{ex-chareq}
with $\hat C^t=\Psi^t$ and $\hat C^x=\Psi^x$,
where the resulting multipliers $Q_\Psi$
are shown in table~\ref{ex-table:multr_symm_adjsymm}.
There is a simple relationship \eqref{multr-symm-multr}
between each multiplier $Q_\Psi$
and the adjoint-symmetry/symmetry pair $Q,P$
used to generate the conserved current $(\Psi^t,\Psi^x)$.
In particular,
from expressions \eqref{ex-adjsymms}--\eqref{ex-adjsymms-R}
for the symmetry characteristics, adjoint-symmetries,
and their associated operators $R_P$ and $R_Q$,
the relationship \eqref{multr-symm-multr} yields
\begin{equation}\label{ex-multr-symm-multr}
\begin{aligned}
& Q_{\Psi(P_1,Q_l;F)} = R_{P_1}^*(Q_l)  - R_{Q_l}^*(P_1) = D_t Q_l,
\quad
l=1,2,3
\\
& Q_{\Psi(P_2,Q_l;F)}= R_{P_2}^*(Q_l)  - R_{Q_l}^*(P_2) = D_x Q_l,
\quad
l=1,2,3
\end{aligned}
\end{equation}
in accordance with table~\ref{ex-table:multr_symm_adjsymm}.

\begin{table}[htb]
\centering
\caption{Multipliers from the adjoint-symmetry/symmetry formula}
\label{ex-table:multr_symm_adjsymm}
\begin{tabular}{c|c||c|c}
\hline
$P$ & $Q$ & $\Psi^t,\Psi^x$ & $Q_\Psi$
\\
\hline
\hline
$ -u_t$
&
$e^{m_3x+m_2t}$
&
$\begin{aligned}
& m_2e^{m_3x+m_2t}(u_t-m_2u+\tint b\,du),
\\
& m_2e^{m_3x+m_2t}(m_3u-u_x+\tint c\,du)
\end{aligned}$
&
$\begin{aligned}
& m_2 e^{m_3x+m_2t} \\&
=D_t(e^{m_3x+m_2t})
\end{aligned}$
\\
\hline
$ -u_x$
&
$e^{m_3x+m_2t}$
&
$\begin{aligned}
& m_3e^{m_3x+m_2t}(u_t-m_2u+\tint b\,du),
\\
& m_3e^{m_3x+m_2t}(m_3u-u_x+\tint c\,du)
\end{aligned}$
&
$\begin{aligned}
& m_3 e^{m_3x+m_2t} \\&
=D_x(e^{m_3x+m_2t})
\end{aligned}$
\\
\hline
\hline
$ -u_t$
&
$e^{\alpha x + \beta t}$
&
$\begin{aligned}
& \beta e^{\alpha x + \beta t}(u_t-\beta u +\tint b\,du),
\\
& \beta e^{\alpha x + \beta t}(\alpha u-u_x+\tint c\,du)
\end{aligned}$
&
$\begin{aligned}
& \beta e^{\alpha x + \beta t} \\&
=D_t(e^{\alpha x + \beta t})
\end{aligned}$
\\
\hline
$ -u_x$
&
$e^{\alpha x + \beta t}$
&
$\begin{aligned}
& \alpha e^{\alpha x + \beta t}(u_t-\beta u +\tint b\,du),
\\
& \alpha e^{\alpha x + \beta t}( \alpha u -u_x +\tint c\,du)
\end{aligned}$
&
$\begin{aligned}
& \alpha e^{\alpha x + \beta t} \\&
=D_x(e^{\alpha x + \beta t})
\end{aligned}$
\\
\hline
\hline
$-u_t$
&
$e^{\gamma x} q$
&
$\begin{aligned}
& -e^{\gamma x}( q'' u \pm q'(u_t +\tint b\,du) ),
\\
& \pm e^{\gamma x}( (\gamma q'-q'')u  + q(u_x \mp\tint b\,du) )
\end{aligned}$
&
$\begin{aligned}
& \mp e^{\gamma x} q \\&
= D_t(e^{\gamma x} q)
\end{aligned}$
\\
\hline
$-u_x$
&
$e^{\gamma x} q$
&
$\begin{aligned}
& e^{\gamma x}( \pm (q'' +\gamma q')u +(q'+\gamma q)(u_t +\tint b\,du) ),
\\
& e^{\gamma x}( (q'' -\gamma^2 q)u -(q'+\gamma q)(u_x\mp\tint b\,du) )
\end{aligned}$
&
$\begin{aligned}
& e^{\gamma x}(q'+\gamma q) \\&
= D_x(e^{\gamma x} q)
\end{aligned}$
\\
\hline
\end{tabular}
\end{table}

In particular, consider the case when $m(u)$ is zero and both $b(u),c(u)$ are arbitrary, 
so then the only admitted multiplier of lowest-order form $Q(t,x,u)$ is $Q=1$ 
(up to a multiplicative constant), 
as shown by table~\ref{ex-table:adjsymm}. 
In this case, it is straightforward to show that (by solving the relevant determining equations) 
there are no first-order multipliers 
and that the only admitted point symmetries are
generated by the translations \eqref{ex-trans-symms}. 
Consequently, when the set of multipliers $Q(t,x,u,u_t,u_x)$ is considered, 
a single non-trivial conservation law 
$C^t=u_t + \smallint b(u)\,du$, $C^x=-u_x + \smallint c(u)\,du$ 
is admitted by the wave equation $u_{tt} -u_{xx}+ b(u)u_t + c(u)u_x=0$ 
with $b(u)$ and $c(u)$ arbitrary,
whereas all of  the conserved currents $(\Psi^t,\Psi^x)$
obtained from Ibragimov's theorem \eqref{ex-L-conslaw},
or from the simpler equivalent adjoint-symmetry/symmetry formula \eqref{ex-conslaw},
are trivial!
Note that, correspondingly, 
the symmetry action on the set of non-trivial conservation laws 
given by the set of multipliers $Q(t,x,u,u_t,u_x)$ is not transitive. 
This example succinctly illustrates the incompleteness of these formulas 
for generating conservation laws.

\section{A direct construction method to find all local conservation laws}
\label{sec:method}

The results stated in Proposition~\ref{correspondence-no-diffids}
and Theorems~\ref{multrdetsys}, \ref{symmaction}, and~\ref{PQformula}
have been developed in \Ref{AncBlu97,AncBlu02a,AncBlu02b,Anc03,Anc16}
and extended in \Ref{AncKar,Anc-review}.
This collective work provides a simple, algorithmic method to find
\emph{all} local conservation laws for any given system of DEs.
The method is based on the general result that
all local conservation laws arise from multipliers
as given by the solutions of a linear system of determining equations,
where the multipliers are simply adjoint-symmetries
subject to certain Helmholtz-type conditions.

Consequently, all multipliers can be found by either of the two following methods
\cite{AncBlu97}:
(1) directly solve the full determining system for multipliers;
or (2) first, solve the determining equation for adjoint-symmetries,
and next, check which of the adjoint-symmetries satisfy the Helmholtz-type conditions.
The adjoint-symmetry determining equation is simply
the adjoint of the symmetry determining equation,
and hence it can be solved by the standard algorithmic procedure used for
solving the symmetry determining equation
\cite{Olv,1stbook,2ndbook}.
Likewise, the same procedure works equally well for solving the
multiplier determining system.

A natural question is, in practice,
at which differential orders $s\geq 0$ will multipliers or adjoint-symmetries
$Q^a(x,u,\p u,\ldots,\p^s u)$ be found?

One answer is that the same situation arises for symmetries.
Normally,
point symmetries are sought first,
since many DE systems admit point symmetries,
and since relatively fewer DE systems admit contact symmetries or higher-order symmetries.
Indeed, the existence of a sufficiently high-order symmetry is
one main definition of an integrable system \cite{MikShaSok},
as this can indicate the existence of an infinite hierarchy of successively higher-order symmetries.
For multipliers,
the most physically important conserved currents always have a low differential order.
Based on numerous examples,
a concrete definition of a \emph{low-order multiplier} that seems to characterize
these physically important conserved currents,
and distinguishes them from higher-order conserved currents
arising for integrable systems,
has been introduced in recent work \cite{AncKar,Anc-review}.

Another answer is that it is straightforward just to find
all multipliers or adjoint-symmetries with a specified differential order
$s=0,1,2,\ldots$,
going up to any desired maximum finite order.
Moreover, in some situations a standard descent/induction argument
\cite{DuzTsu,AncPoh01,AncPoh03}
can be used to find the multipliers or adjoint-symmetries to all orders $s\geq 0$.

Once a set of multipliers has been found for a given DE system,
the corresponding conserved currents are straightforward to find in an explicit form.
Several different methods are available.

One algorithmic method is the direct integration of the characteristic equation \cite{2ndbook,Wol}
defining the conserved current.
Another algorithmic method is the use of a homotopy integral formula.
This method has several versions \cite{Olv,AncBlu97,AncBlu02a,AncBlu02b,Anc-review},
all of which involve trade-offs between the simplicity of the integration
versus the flexibility of avoiding singularities (if any) in the integrand.

However,
purely algebraic methods for the construction of conserved currents from multipliers
are known.
One algebraic method is the use of a scaling formula \cite{Anc03,Anc16,2ndbook}
which is given by the adjoint-symmetry/symmetry formula.
This applies only to DE systems that admit a scaling symmetry,
but it has recently been extended to general DE systems by
incorporating a dimensional analysis method as shown in \Ref{Anc-review}.
In particular, with the use of this dimensional-scaling method,
the construction of conserved currents becomes completely algebraic.

Therefore,
the general method just outlined
provides a completely algorithmic computational way
to derive all local conservation laws for any given DE system.
In particular,
there is no need to resort to any special methods or ansatzes
(such as
the ``abc'' technique \cite{Mor},
partial Lagrangians \cite{KarMah},
``nonlinear self-adjointness'' \cite{Ibr07,Ibr11,Gan11,Gan14,ZhaXie},
undetermined coefficients \cite{PooHer})
which at best just
yield a subset of all of the local conservation laws admitted by a DE system
or just apply to restricted classes of DE systems.

{\bf Example}:
The semilinear wave equation \eqref{ex-eqn} can be expected to admit
conserved currents that depend nonlinearly on $u_t$ and $u_x$,
in addition to the previous conserved currents in table~\ref{ex-table:conslaw},
all of which have linear dependence on $u_t$ and $u_x$.
The multipliers \eqref{ex-adjsymms} for the latter conserved currents
have the form $Q_C(t,x)$.
Conserved currents that depend nonlinearly on $u_t$ and $u_x$
will arise from multipliers $Q_C(t,x,u,u_t,u_x)$
that have explicit dependence on $u_t$ and $u_x$.
It is straightforward to find all such multipliers by using Maple to set up and solve
the multiplier equation \eqref{ex-multreq},
which splits respect to the variables
$u_{tx},u_{tt},u_{xx},u_{txx},u_{ttx},u_{txx},u_{ttt},u_{xxx}$,
giving an overdetermined linear system.
Alternatively, the multiplier equation \eqref{ex-multreq}
can be split instead into the two terms
$\delta^*_{Q_C} F$ and $\delta^*_F Q_C$,
which provides a direct connection between multipliers and adjoint-symmetries.
In particular,
the first term in the multiplier equation \eqref{ex-multreq} consists of
\begin{equation}\label{ex-1stord-split1}
 \delta^*_{Q_C} F
= D_t^2Q_C  -D_x^2Q_C -b D_tQ_C -c D_x Q_C +(b' u_t+c' u_x+m')Q_C = R_{Q_C}(F)
\end{equation}
where the operator $R_{Q_C}$ is found to be given by
\begin{equation}\label{ex-1stord-R_Q}
\begin{aligned}
R_Q & =
\parder{Q_C}{u_t}D_t F +\parder{Q_C}{u_x}D_x F
+ \parder{{}^2Q_C}{u_t\partial u_t} F
+ \parder{Q_C}{u} - 2b\parder{Q_C}{u_t}
+2u_t\parder{{}^2Q_C}{u\partial u_t}
\\&\qquad
+2u_{tx}\parder{{}^2Q_C}{u_x\partial u_t}
+2(u_{xx}-bu_t-cu_x-d) \parder{{}^2Q_C}{u_t\partial u_t}
\end{aligned}
\end{equation}
for multipliers $Q_C(t,x,u,u_t,u_x)$,
through $u_{tt}=u_{xx} -b(u)u_t -c(u)u_x -m(u)$ and its differential consequences. 
The second term in the multiplier equation \eqref{ex-multreq} is given by
\begin{equation}\label{ex-1stord-split2}
\delta^*_F Q_C
= \parder{Q_C}{u}F -D_t\Big(\parder{Q_C}{u_t}F\Big) -D_x\Big(\parder{Q_C}{u_x}F\Big)
= -\parder{Q_C}{u_t}D_t F -\parder{Q_C}{u_x}D_x F +E_u(Q) F
\end{equation}
where
\begin{equation}\label{ex-1stord-eulerQ}
\begin{aligned}
E_u(Q) & =
- \parder{{}^2Q_C}{u_t\partial u_t} F
+ \parder{Q_C}{u}
-u_t\parder{{}^2Q_C}{u\partial u_t} -u_x\parder{{}^2Q_C}{u\partial u_x}
-2u_{tx}\parder{{}^2Q_C}{u_x\partial u_t}
\\&\qquad
+(bu_t+cu_x+d-2u_{xx})\parder{{}^2Q_C}{u_t\partial u_t}
\end{aligned}
\end{equation}
for multipliers $Q_C(t,x,u,u_t,u_x)$.
On the solution space $\Esp$ of the wave equation \eqref{ex-eqn},
the terms \eqref{ex-1stord-split2} vanish,
while the other terms \eqref{ex-1stord-split1} reduce to the adjoint-symmetry equation \eqref{ex-adjsymmeq}.
Off of the solution space $\Esp$,
the these terms \eqref{ex-1stord-split1} and \eqref{ex-1stord-split2}
become a linear combination of $F$, $D_t F$, $D_x F$,
whose coefficients must vanish separately.
This splitting is found to yield a single Helmholtz-type equation
\begin{equation}\label{ex-1stord-helmholtzeq}
2\parder{Q_C}{u}
-b\parder{Q_C}{u_t}
+u_t\parder{{}^2Q_C}{u\partial u_t} -u_x\parder{{}^2Q_C}{u\partial u_x}
-(bu_t +cu_x +d) \parder{{}^2Q_C}{u_t\partial u_t}
=0 .
\end{equation}
Taken together,
this Helmholtz-type equation \eqref{ex-1stord-helmholtzeq}
and the adjoint-symmetry equation \eqref{ex-adjsymmeq}
constitute the determining system \eqref{multr-adjsymmeq}--\eqref{multr-helmholtzeq}
for finding all first-order multipliers $Q_C(t,x,u,u_t,u_x)$
admitted by the wave equation \eqref{ex-eqn}.

The most computationally effective way to solve equations
\eqref{ex-1stord-helmholtzeq} and  \eqref{ex-adjsymmeq} in the determining system
is by changing variables from $t,x,u,u_t,u_x$ to
$\mu=\tfrac{1}{2}(t+x)$, $\nu=\tfrac{1}{2}(t-x)$, $u$, $u_\mu = u_t+u_x$, $u_\nu = u_t-u_x$,
based on null coordinates for the wave equation \eqref{ex-eqn}.
In these new variables,
the general solution of the determining system consists of
three distinct cases (as obtained using the Maple package 'rifsimp'),
after the nonlinearity and homogeneity conditions \eqref{ex-conds} are imposed
on $b(u),c(u),d(u)$.
The resulting multipliers, after merging cases,
are shown in table~\ref{ex-table:multr-1stord}.
Each multiplier determines a non-trivial conserved current
through the characteristic equation \eqref{ex-chareq}.
These conserved currents $(\hat C^t,\hat C^x)$ can be derived
in terms of the multipliers $Q_C(t,x,u,u_t,u_x)$
in the same way discussed previously for lowest-order multipliers.
The results are shown in table~\ref{ex-table:conslaw-1stord}.

\begin{table}[htb]
\centering
\caption{First-order multipliers}
\label{ex-table:multr-1stord}
\begin{tabular}{c|c}
\hline
conditions & $Q_C$
\\
\hline
\hline
$\begin{aligned}
& \frac{2m_1m_2}{m-m_1} = \frac{4m_1}{b\pm c} = \tint (b\mp c)\,du
\end{aligned}$
&
$\dfrac{2m_1+(b\pm c)(u_t\pm u_x)}{2m +(b\pm c)(u_t\pm u_x)}$
\\
\hline
$\begin{aligned}
& m= (m_1 + \tfrac{1}{4}\tint (b-c)\,du)(b+c),
\\
& (1-\gamma)b = (1+\gamma)c
\end{aligned}$
&
$\dfrac{((1-\gamma)u_t +(1+\gamma)u_x)(b^2-c^2)}{((b+c)(u_t+u_x)+2m)((b-c)(u_t-u_x)2m)}$
\\
\hline
\end{tabular}
\end{table}

\begin{table}[htb]
\centering
\caption{First-order conserved currents}
\label{ex-table:conslaw-1stord}
\begin{tabular}{c|c}
\hline
conditions & $\hat C^t$, $\hat C^x$
\\
\hline
\hline
$\begin{aligned}
& \frac{2m_1m_2}{m-m_1} = \frac{4m_1}{b\pm c} = \tint (b\mp c)\,du
\end{aligned}$
&
$\begin{aligned}
& \gamma\ln\Big(\frac{b\pm c}{2m_1 +(b\pm c)(\gamma+u_t\pm u_x)}\Big)
+u_t+ \tfrac{1}{2}\tint(b\pm c)\,du,
\\
& \mp\gamma\ln\Big(\frac{b\pm c}{2m_1 +(b\pm c)(\gamma+u_t\pm u_x)}\Big)
-u_x + \tfrac{1}{2}\tint(c\pm b)\,du +\gamma x
\end{aligned}$
\\
\hline
$\begin{aligned}
& m= \big(m_1 + \tfrac{1}{4}\tint (b-c)\,du\big)(b+c),
\\
& (1-\gamma)b = (1+\gamma)c
\end{aligned}$
&
$\begin{aligned}
& \ln\bigg(\frac{\big(\gamma\big( \tint(b+ c)\,du +2(u_t-u_x)\big) +m_1\big)^{\frac{1}{\gamma}}}{\gamma \tint(b+ c)\,du +2(u_t+u_x) +m_1}\bigg),
\\
& \ln\Big(\big(\gamma\big( \tint(b+ c)\,du +2(u_t-u_x)\big) +m_1\big)^{\frac{1}{\gamma}}
\\&\qquad
\times \big(\gamma \tint(b+ c)\,du +2(u_t+u_x) +m_1\big)\Big)
\end{aligned}$
\\
\hline
\end{tabular}
\end{table}

\section{Concluding remarks}
\label{sec:remarks}

The conservation law theorem stated by Ibragimov in \Ref{Ibr07,Ibr11}
for ``nonlinear self-adjoint'' DEs
and subsequent extensions of this theorem in \Ref{Gan11,Gan14,ZhaXie} are not new.
In its most general form,
this theorem is simply a re-writing of a standard formula \cite{Cav,Lun,AncBlu97}
that uses a pair consisting of a symmetry and an adjoint-symmetry
to produce a conservation law through a well-known Fr\'echet  derivative identity \cite{Cav,Lun,Olv,2ndbook,Anc-review}.
Unfortunately, no references to prior literature are provided in Ibragimov's papers,
which may give the impression that the results are original.
One aspect that is novel is the derivation of the formula
by using an auxiliary Lagrangian,
although it does not in any way simplify either the formula or its content.
Moreover, the condition of ``nonlinear self-adjointness'' is nothing but a re-writing of
the condition that a DE system admits an adjoint-symmetry \cite{AncBlu97,Anc-review},
and this condition automatically holds for any DE system that admits a local conservation law.

The present paper shows how the symmetry/adjoint-symmetry formula
is directly connected to the action of symmetries on conservation laws,
which explains a number of major drawbacks in trying to use the formula
--- and hence in applying Ibragimov's theorem ---
as a method to generate conservation laws.
In particular, the formula can generate trivial conservation laws
and does not always yield all non-trivial conservation laws
unless the symmetry action on the set of these conservation laws is transitive,
which cannot be known until all conservation laws have been found.

A broader point, which is more important, is that
there is a completely general method \cite{2ndbook,Anc-review}
using adjoint-symmetries \cite{Cav,Lun,AncBlu97,AncBlu02a,AncBlu02b}
to find all local conservation laws for any given DE system.
This method is a kind of adjoint version of the standard Lie method to find all local symmetries.
The method is algorithmic \cite{Anc-review}
and the required computations are no more difficult
than the computations used to find local symmetries.

\section*{Acknowledgements}
The author is supported by an NSERC Discovery grant.
M.\ Gandarias is thanked for helpful discussions.


\begin{thebibliography}{99}

\bibitem{Noe}
E. Noether,
Invariante Variationsprobleme,
{\em Nach. Konig. Gesell. Wissen. Gottingen, Math.-phys. Kl}
(1918), 235--257;
English translation in {\em Transport Theor. and Stat. Phys.} 1 (1971), 186--207.

\bibitem{Cav}
G. Caviglia,
Symmetry transformations, isovectors, and conservation laws,
{\em J. Math. Phys.} 27 (1986), 972--978.

\bibitem{Lun}
F. A. Lunev, 
An analogue of the Noether theorem for non-Noether and nonlocal symmetries. (Russian) {\em Teoret. Mat. Fiz.} 84 (1990), no. 2, 205--210; 
translation in {\em Theoret. and Math. Phys.} 84 (1990), no. 2, 816–-820. 

\bibitem{AncBlu97}
S.C. Anco and G. Bluman,
Direct construction of conservation laws from field equations,
{\em Phys. Rev. Lett.} 78 (1997), 2869--2873.

\bibitem{Bla}
M. Blaszak, 
{\em Multi-Hamiltonian theory of dynamical systems}
Springer:Heildelberg, 1998.

\bibitem{KraVin}
I.S. Krasil'shchik and A.M. Vinogradov (eds.),
{\em Symmetries and Conservation Laws for Differential Equations of Mathematical Physics} ,
Translations of Mathematical Monographs 182,
Amer. Math. Soc.:Providence, 1999.

\bibitem{Ver}
A. Verbotevsky,
Notes on the horizontal cohomology, 
in {\em Secondary Calculus and Cohomological Physics}, 211--232,
Contemporary Mathematics 219,
Amer. Math. Soc.:Providence, 1997.

\bibitem{Zha86}
V.V. Zharinov, 
{\em Lecture notes on geometrical aspects of partial differential equations}, 
Series on Soviet and East European Mathematics, Vol. 9., 
World Scientific: River Edge, 1992. 

\bibitem{Ibr07}
N.H. Ibragimov,
A new conservation theorem,
{\em J. Math. Anal. Appl.} 333 (2007), 311--328.

\bibitem{Ibr07b}N.H. Ibragimov, 
Quasi self-adjoint differential equations, 
{\em  Arch. ALGA} 4 (2007), 55--60. 

\bibitem{Ibr10}
N.H. Ibragimov,
Nonlinear self-adjointness in constructing conservation laws,
{\em Archives of ALGA} 7/8 (2010), 1--86.

\bibitem{Ibr11}
N.H. Ibragimov,
Nonlinear self-adjointness and conservation laws,
{\em J. Phys. A: Math. Theor.} 44 (2011), 432002 (8pp).

\bibitem{Gan11}
M.L. Gandarias,
Weak self-adjoint differential equations,
{\em J. Phys. A: Math. Theor.} 44 (2011), 262001 (6pp).

\bibitem{GalIbr}
L.R. Galiakberova, N.H. Ibragimov, 
Nonlinear self-adjointness of the KricheverNovikov equation,
{\em Commun. Nonlinear Sci. Numer. Simul.} 19 (2014),  361--363. 

\bibitem{Gan14}
M.L. Gandarias,
Nonlinear self-adjointness through differential substitutions,
{\em Commun. Nonlinear Sci. Numer. Simul.} 19 (2014),  3523--3528.

\bibitem{Zha}
Z.-Y. Zhang,
On the existence of conservation law multiplier for partial differential equations,
{\em Commun. Nonlinear Sci. Numer. Simul.} 20 (2015), 338--351.

\bibitem{ZhaXie}
Z.-Y. Zhang, L. Xie,
Adjoint symmetry and conservation law of nonlinear diffusion equations with convection and source terms,
{\em Nonlinear Analysis: Real World Applications} 32 (2016) 301--313.

\bibitem{Fre13}
I.L. Freire,
New classes of nonlinearly self-adjoint evolution equations of third- and fifth-order,
{\em Commun. Nonlin. Sci. Numer. Simul.} 18 (2013), 493--499.

\bibitem{Gan15}
M.L. Gandarias,
Conservation laws for some equations that admit compacton solutions induced by a non-convex convection,
{\em J. Math. Anal. Appl.} 420 (2015) 695--702.

\bibitem{IbrTorTra10}
N. H. Ibragimov, M. Torrisi, R. Tracina,
Quasi self-adjoint nonlinear wave equations,
{\em J. Phys. A: Math. Theor.} 43 (2010) 442001 (8pp).

\bibitem{IbrTorTra11}
N. H. Ibragimov, M. Torrisi, R. Tracina,
Self-adjointness and conservation laws of a generalized Burgers equation,
{\em J. Phys. A: Math. Theor.} 44 (2011) 145201 (5pp).

\bibitem{FreSam12}
I.L. Freire, J.C.S. Sampaio,
Nonlinear self-adjointness of a generalized fifth-order KdV equation,
{\em J. Phys. A: Math. Theor.} 44 (2012) 032001 (7pp).

\bibitem{Anc16}
S.C. Anco,
Symmetry properties of conservation laws,
{\em Int. J. Mod.  Phys. B} 30 (2016), 1640004 (12pp).

\bibitem{Olv}
P. Olver,
{\em Applications of Lie Groups to Differential Equations},
Springer-Verlag, New York, 1986.

\bibitem{Mar-Alo}
L. Martinez Alonso,
On the Noether map,
{\em Lett. Math. Phys.} 3 (1979), 419--424.

\bibitem{AncBlu02a}
S.C. Anco and G. Bluman,
Direct construction method for conservation laws of partial differential equations.~
I. Examples of conservation law classifications,
{\em Euro. J. Appl. Math.} 13 (2002), 545--566.

\bibitem{AncBlu02b}
S.C. Anco and G. Bluman,
Direct construction method for conservation laws of partial differential equations.~
II. General treatment,
{\em Euro. J. Appl. Math.} 13 (2002), 567--585.

\bibitem{Anc03}
S.C. Anco,
Conservation laws of scaling-invariant field equations, 
{\em J. Phys. A: Math. Gen.} 36 (2003), 8623--8638.

\bibitem{Anc-review}
S.C. Anco,
Generalization of Noether's theorem in modern form to non-variational partial differential equations, 
in {\em Recent progress and Modern Challenges in Applied Mathematics, Modeling and Computational Science}, 
Fields Institute Communications, Vol. 79, 2017. 

\bibitem{CRC}
{\em CRC Handbook of Lie Group Analysis of Differential Equations, Volume I: Symmetries, Exact Solutions, and Conservation Laws} (ed. N.H. Ibragimov), 
CRC Press: Boca Raton, 1994. 

\bibitem{2ndbook}
G. Bluman, A. Cheviakov, S.C. Anco,
{\em Applications of Symmetry Methods to Partial Differential Equations},
Springer Applied Mathematics Series 168,
Springer, New York, 2010.

\bibitem{1stbook}
G. Bluman and S.C. Anco,
{\em Symmetry and Integration Methods for Differential Equations},
Springer Applied Mathematics Series 154,
Springer-Verlag, New York, 2002.

\bibitem{Kha}
R.S. Khamitova, 
The structure of a group and the basis of conservation laws. (Russian) 
{\em Teoret. Mat. Fiz.} 52 (1982), no. 2, 244-–251;
translation in {\em Theoret. and Math. Phys.} 52, (1982), no. 2, 777-–781. 

\bibitem{Anc17}
S.C. Anco,
On the algebra of symmetries and adjoint-symmetries of differential equations,
in preparation (2017).

\bibitem{AncBlu96}
S.C. Anco and G. Bluman,
Derivation of conservation laws from nonlocal symmetries of differential equations,
{\em J. Math. Phys.} 37 (1996), 2361--2375.

\bibitem{AncKar}
S.C. Anco and A.H. Kara,
Symmetry-invariant conservation laws of partial differential equations, 
In press {\em Euro. J. Appl. Math.}, (2017). 

\bibitem{MikShaSok}
A.V. Mikhailov, A.B. Shabat and V.V. Sokolov,
Symmetry approach to classification of integrable equations,
in {\em What is Integrability?}
(ed. V.E. Zakharov), 
Springer-Verlag, 1999.

\bibitem{DuzTsu}
S.V. Duzhin and T. Tsujishita,
Conservation laws of the BBM equation,
{\em J. Phys. A} 17 (1984), 3267--3276.

\bibitem{AncPoh01}
S.C. Anco and J. Pohjanpelto,
Classification of local conservation laws of Maxwell's equations,
{\em Acta. Appl. Math.} 69 (2001), 285--327.

\bibitem{AncPoh03}
S.C. Anco and J. Pohjanpelto,
Conserved currents of massless spin s fields,
{\em Proc. Roy. Soc.} 459 (2003), 1215--1239.

\bibitem{Wol}
T. Wolf,
A comparison of four approaches to the calculation of conservation laws,
{\em Euro. J. Appl. Math.} 13 (2002), 129--152.

\bibitem{Mor}
C. Morawetz,
Variations on conservation laws for the wave equation,
{\em Bulletin Amer. Math. Soc.} 37 (2000), no. 2, 141--154.

\bibitem{KarMah}
A.H. Kara and F.M. Mahomed,
Noether-type symmetries and conservation laws via partial Lagrangians,
{\em Nonlin. Dyn.} 45 (2006), 367--383.

\bibitem{PooHer}
D. Poole and W. Hereman,
Symbolic computation of conservation laws for nonlinear partial differential equations in multiple space dimensions,
{\em J. Symbolic Computation} 46 (2011), 1355--1377.

\end{thebibliography}
\end{document}